\documentclass[%
 reprint,
 twocolumn,
amsmath,
 amssymb,
 aps,
prb,
]{revtex4-2}

\usepackage{dcolumn}
\usepackage{bm}
\usepackage[english]{babel}
\usepackage[utf8]{inputenc}
\usepackage[colorinlistoftodos, color=green!40, prependcaption]{todonotes}

\usepackage{amsthm}
\usepackage{mathtools}
\usepackage{physics}
\usepackage{xcolor}
\usepackage{graphicx}

\usepackage{bbold}

\usepackage{adjustbox}
\usepackage{placeins}

\usepackage{csquotes}
\newtheorem{theorem}{Theorem}

\usepackage[caption=false]{subfig}
 \usepackage[colorlinks=true,linkcolor=blue]{hyperref}
 
 \makeatletter
\renewcommand{\fnum@figure}{FIG. \thefigure}
\makeatother
\usepackage{tikz}
\usetikzlibrary{decorations.markings}

\usepackage{amsmath,mleftright}
\usepackage{xparse}

\NewDocumentCommand{\evalat}{sO{\big}mm}{%
  \IfBooleanTF{#1}
   {\mleft. #3 \mright|_{#4}}
   {#3#2|_{#4}}%
}

\begin{document}

\preprint{APS/123-QED}

\title{
Unified formulation of interfacial magnonic pumping from non-collinear magnets}

\author{Virgile Guemard}
\email{virgile.guemard@gmail.com}
\author{Aurélien Manchon}%
 \email{manchon@cinam.univ-mrs.fr}
\affiliation{CINaM, Aix-Marseille Univ, CNRS, Marseille, France
}%




\date{\today}

\begin{abstract}
We establish a general formulation for spin current pumped by magnons at the interface between a normal metal and a magnetic insulator, valid for any non-collinear magnetic configuration in the linear spin wave regime. This current is generated by driving the system in a non-equilibrium state, covering setups such as thermal spin injection (spin Seebeck effect) or spin voltage by irradiation  of the insulator (spin pumping). We illustrate the formula in the special case of a honeycomb topological ferromagnet, for simplicity, and cover both the spin Seebeck and spin-pumping setups. We show how the topological parameters influence the spin current and propose a way to obtain a contribution mainly from the topological edge magnons in the spin pumping case.

 \end{abstract}

\maketitle


\section{Introduction}

Transport of magnons in insulators as well as their ability to be pumped through other materials are of great interest due to absence of Joule heating and their functionality for high speed data transport at acceptable temperature \cite{Chumak2015}. In addition, the same exotic properties displayed by conduction electrons in metals, namely the various flavors of the Hall effects including topological states at their boundaries, have recently been experimentally reported in magnonic materials \cite{Onose2010}, drawing a net analogy with electronic topological Chern insulators \cite{Haldane1988}. The existence of magnonic edge channels naturally leads to the proposal of several applications based on coherent magnon transport, such as magnon beam splitters or magnonic Fabry-Perot interferometer \cite{Shindou2013,Wang2017b,Wang2018}. These breakthroughs in the topological realm have boosted the interest for magnon transport in magnetic insulators, paving the way for the exploitation of magnons as quantum information carriers. Consequently extensive efforts are achieved towards the identification of topological phases of magnon materials and it has been shown that magnonic topological phases can also exist in non-collinear magnets \cite{Mook2019}, opening prospects to a vast new panel of topological materials. Despite this intense effort and the promises it bears, most works have been limited to topological band structure characterization. Such a characterization often boils down to Chern number and edge state computations, which are hardly accessible under experimental conditions. In this Article, we suggest that topological magnonic egde states are in fact accessible via interfacial magnonic spin pumping, as usually achieved in spin caloritronics experiments \cite{Uchida2010,Xiao2010,Bender2012,Zhang2012}. To achieve this goal and identify the signature of interfacial magnons on the pumped spin current, we theoretically derive a general formula for spin pumping current adapted to both collinear and noncollinear magnets. We then apply it to a topological ferromagnet and characterize the contribution of topological states to the spin current response.\\ 

The issue we address in this paper is twofold. Spin injection theories at the interface of a heterostructure between a magnet and a metal  under a temperature gradient (spin-Seebeck effect) or driven by ferromagnetic resonance
(FMR) \cite{Tserkovnyak2002,Ando2008,Kajiwara2010,Patra2012,RojasSnchez2013, Hahn2013,Saitoh2006} (spin pumping) have been successfully established in the past \cite{Uchida2008,Adachi2013,Ohnuma2014,Kamra2017,Matsuo2018,Kato2019}. However these approaches are limited to collinear ferro-, ferri- and antiferromagnets, disregarding non-collinear systems which in fact host a wealth of unconventional magnetic and electronic properties \cite{Bonbien2021}. Consequently, the literature on spin-transport involving non-collinear magnets is still scarce and the issue of spin pumping between non-collinear magnets and nonmagnetic metals has only been treated in recent works \cite{Flebus2019,Ma2020}. We here propose a compact formula in a Landauer form \cite{Landauer1957}, derived microscopically \cite{Caroli1971} with the most general lattice model, formulated in terms of the local spin susceptibility, i.e the spin density correlation function. 
\\

Second, the spin current pumped out of a topological magnonic insulator into an adjacent metal is expected to display specific properties, associated with the magnon concentration at the interface. The theories developed to date are limited to \emph{bulk} magnons and overlook the role of edge states. These states being topologically protected, they weakly interact with bulk magnons and are therefore expected to display unique signature in the spin current. A major hurdle lies in the fact that magnons are bosonic particles so that the whole band structure contributes to the transport at finite temperature. A way to discriminate the contribution of edge and bulk magnons is thus needed.\\

The present paper is organized as follows. Section~\ref{sec:Model} contains a presentation of the Hamiltonian model and introduces the linear spin wave theory notations. In Section~\ref{sec: dynamical spin susceptibility}, the dynamical spin susceptibilities are defined and the spin-transport formalism is established. The computation of the spin-current is the object of Section~\ref{sec: spin-current}. The theoretical methodology is based on Keldysh formalism \cite{Rammer1986} and non-restrictive assumptions are made along the way to give a compact and readily usable Landauer formula. Section~\ref{sec: application} applies our formalism to a topological magnon insulator introduced in \cite{Owerre2016}. Concluding remarks are given in Section~\ref{sec: conclusion}.

\section{model}\label{sec:Model}

\begin{figure}[t]
\includegraphics[width=0.47\textwidth]{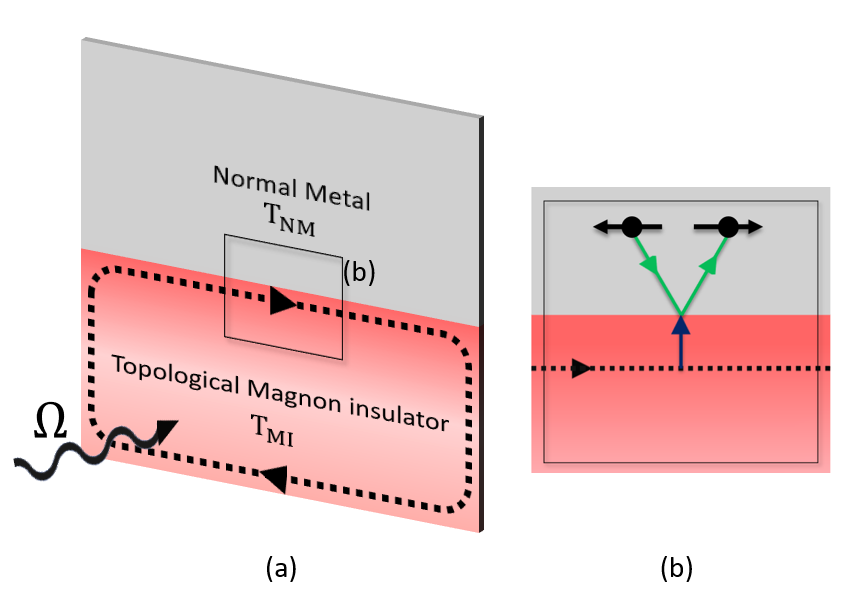}
\caption{\label{fig:homemade}(a) Schematic picture of the magnetic insulator/nonmagnetic metal bilayer under a temperature gradient ($T_{\rm MI}>T_{\rm NM}$) or FMR drive of the magnetic insulator at frequency $\Omega$, responsible for spin injection from the insulator into the metal. The insulator is in its topological phase and magnon edge modes propagate. (b) Example of mechanism for spin injection. The exchange interaction at the interface of the bilayer induces electron spins to flip as they reflect on the edge. For simplicity the precession of the magnetic spins are shown in the same direction as their propagation.}
\end{figure}

In this article, we consider a bilayer composed of a magnetic insulator adjacent to a nonmagnetic metal, as usually encountered in spin pumping and spin caloritronic experiments \cite{Saitoh2006, Uchida2010}. The detailed formalism for dealing with magnons in non-collinear magnets has been exposed in detailed in Refs. \cite{Colpa1978,Shindou2013,Mook2019}. For the sake of completeness, we summarize the complete procedure in this section and invite the readers familiar with this approach to directly move to Section. \ref{sec: dynamical spin susceptibility}. The Hamiltonian under consideration can always be decomposed in four parts,
\begin{equation}
H=H_{\rm MI}+H_{\rm NM}+H_{\rm int}+ V.
\end{equation}
The first term corresponds to a general magnetic insulator (MI). The second term models the normal metal (NM) adjacent to the magnetic insulator and isolated from it. The third term contains the interaction between the two materials (see Fig. \ref{fig:homemade}). The last term is an external perturbation acting on the insulator. Its value is $V=0$ for spin-Seebeck effect setup and $V\ne 0$ when an external magnetic field irradiates the insulator to drive FMR.

\subsection{Metallic system}

The normal metal is treated as an ideal spin-sink: spins relax sufficiently fast at its interface so that there is no spin accumulation. We assume that it is quadratic in fermionic operators and spin mixing is forbidden (in other words, spin-orbit coupling or non-collinear magnetism are absent). Therefore the electronic Hamiltonian simply reads
\begin{equation}
H_{\rm NM} = \sum_{\mathbf{q}\mathbf{q}',\sigma} c^{\dagger}_{\mathbf{q},\sigma}[ \epsilon_\mathbf{q}\delta_{\mathbf{q} \mathbf{q}'}+U_{\mathbf{q} -\mathbf{q}'}] c_{\mathbf{q}',\sigma},
\end{equation}
$U_{\mathbf{q} -\mathbf{q}'}$ being the Fourier transform of an impurity potential, and $c^{(\dagger)}_{\mathbf{q},\sigma}$ the fermionic operators of spin $\sigma$ and momentum ${\bf q}$. One can define the local spin density as $
\textbf{s}({\bf R}) =  c^{\dagger}({\bf R}) \boldsymbol\sigma c({\bf R})$ and express it in momentum space
\begin{equation}
\textbf{s}_{\mathbf{q}} = \frac{1}{\sqrt{N}} \sum_{\mathbf{p}} c^{\dagger}_{\mathbf{p+q}} \boldsymbol\sigma c_{\mathbf{p}},
\end{equation}
with $c_{\mathbf{q}}=(c_{\mathbf{q},\sigma=\uparrow},c_{\mathbf{q},\sigma=\downarrow})^T$ and $\boldsymbol\sigma$ being the vector of Pauli matrices.

\subsection{Magnetic system}\label{Colpa}
The magnetic insulator is taken as general as possible in order to describe magnets possessing collinear or non-collinear configurations. The Hamiltonian is supposed to be quadratic in the spin operator and can be written in full generality
\begin{equation}\label{eq: magnetic system local}
H_{\rm MI}=\sum_{i,j}\mathbf{S}_i^T\mathcal{J}_{ij}\mathbf{S}_j,
\end{equation}
where $i,j$ indices describe the position of a spin in the lattice, and $\mathcal{J}_{ij}$ is the interaction matrix.

An efficient procedure to diagonalize Eq. \eqref{eq: magnetic system local} starts by assuming that it respects the ansatz of Linear Spin Wave Theory which treats small fluctuations of spin around their equilibrium positions. This assumption enables us to treat these quanta as bosonic modes of excitation and hence map a spin Hamiltonian into a bosonic one. The method starts by evaluating the ground state configuration of this system, considering quantum mechanical spins as classical vectors oriented in 3D space.
The orientation of each classical spin-vector is then known and an ordering into magnetic cells should appear. The Hamiltonian can be expressed as
 \begin{equation}
H_{\rm MI}=\sum_{l,m}^M\sum_{i,j}^N\mathbf{S}_{l,i}^T\mathcal{J}_{ij}^{lm}\mathbf{S}_{m,j},
\end{equation}
where now each spin is expressed as $\textbf{S}_{m,i}$, the first index corresponding to the $m$th magnetic cell ($m=1,...,M$) and the second index representing the position $i$ ($i=1,...,N$) of the ion within the cell. The position of $\textbf{S}_{m,i}$ is noted $\textbf{R}_m+\textbf{R}_i$. A spin can be connected to the global frame of the system using a rotation
\begin{equation}
\textbf{S}_{m,i}=\mathcal{R}_z(\phi_i)\mathcal{R}_y(\theta_i)\textbf{S'}_{m,i}=\mathcal{R}_i \textbf{S'}_{m,i},
\end{equation}
$\textbf{S'}_{m,i}$ being the spin operator in its own local frame. It means that if we consider the ground state (no magnons), $\textbf{S'}_{m,i}=(0,0,S)^T$ for all $m$ and $i$. $\mathcal{R}_i$ depends only on sites $i$ within the magnetic cell and can be written
\begin{equation}
\mathcal{R}_i=\begin{pmatrix}
\cos \theta_i \cos\phi_i & -\sin\phi_i & \sin \theta_i \cos\phi_i\\
\cos \theta_i \sin\phi_i & \cos \phi_i & \sin \theta_i \sin\phi_i\\
-\sin \theta_i & 0& \cos\theta_i
\end{pmatrix}
=(\hat{\textbf{x}}_i,\hat{\textbf{y}}_i,\hat{\textbf{z}}_i).
\end{equation}
In their local frame, each spin operator can be transformed using a Holstein-Primakoff \cite{Holstein1940} transformation in the large $S$ limit,
\begin{equation}\label{holstein}
\textbf{S'}_{m,i}=\begin{pmatrix}\sqrt{S_i/2}(b_{m,i}+b^\dagger_{m,i})\\
-i\sqrt{S_{i/2}}(b_{m,i}-b^\dagger_{m,i})\\
S_i - b^\dagger_{m,i} b_{m,i}
    \end{pmatrix},
\end{equation}
where $b^{(\dagger)}_{m,i}$ are bosonic annihilation (creation) operators in real space. In the global reference frame it yields,

\begin{equation}\label{Smi}
\textbf{S}_{m,i}=\sqrt{S_i/2}(b^{\dagger}_{m,i} \hat{\textbf{u}}_i +b_{m,i} \hat{\textbf{u}}^*_i )  + (S_i - b^{\dagger}_{m,i}b_{m,i})\hat{\textbf{z}}_i,
\end{equation}
with $\hat{\textbf{u}}^{(*)}_i= \hat{\textbf{x}}_i \pm i\hat{\textbf{y}}_i$. In momentum space these bosonic operators are expressed
\begin{equation}
b^{(\dagger)}_{m,i}=\frac{1}{\sqrt{N}}\sum_{\textbf{k}}e^{i \textbf{k}.\textbf{R}_m}b^{(\dagger)}_{(-)\textbf{k},i}.
\end{equation}
It is then straightforward to express a quadratic Hamiltonian in the Bogoliubov-de-Gennes (BdG) form,

\begin{equation}\label{Nambu}
H_{\rm MI}=\frac{1}{2}\sum_k \textbf{b}^\dagger_\textbf{k}\mathcal{H}_\textbf{k} \textbf{b}_\textbf{k},
\end{equation}
with $\textbf{b}$ a column vector of bosonic operators,

\begin{equation}\label{spinor}
\textbf{b}^\dagger_\textbf{k}=(b^\dagger_{\textbf{k},1},...,b^\dagger_{\textbf{k},N},b_{\textbf{-k},1},...,b_{\textbf{-k},N}),
\end{equation}
satisfying the commutation relation
\begin{equation}
[\textbf{b}_\textbf{k},\textbf{b}^\dagger_\textbf{k}]=\eta,
\end{equation}
with
\begin{equation}
    \eta= \begin{pmatrix}
\mathbb{1}_N & 0 \\
0 & -\mathbb{1}_N 
\end{pmatrix}.
\end{equation}
 A general procedure to diagonalize the Hamiltonian is proposed in \cite{Colpa1978} and we give now a pedagogical summary. First of all, because of the construction of Eq. (\ref{spinor}), the blocks of $\mathcal{H}_k$ are not independent of each others. It is required to be of the following form
\begin{equation}
    \mathcal{H}_k= \begin{pmatrix}
A_\mathbf{k} & B_\mathbf{k} \\
B^*_{\mathbf{-k}} & A^*_{\mathbf{-k}} 
\end{pmatrix},
\end{equation}
in order to fulfill hermiticity. It is then diagonalized by a Bogoliubov-Valatin (BV) transformation. We recall that a BV transformation is an invertible linear transformation of creation and annihilation operators preserving their algebraic relations. It is not required to be unitary but symplectic for bosons. Writing this transformation $P_\textbf{k}$, we have
\begin{equation}
[\textbf{a}_\textbf{k},\textbf{a}^\dagger_\textbf{k}]=[P_\textbf{k}^{-1}\textbf{b}_\textbf{k},(P_\textbf{k}^{-1}\textbf{b}_\textbf{k})^\dagger]=\eta,
\end{equation}
and
\begin{equation}\label{Nambu}
 \textbf{b}^\dagger_\textbf{k}\mathcal{H}_k \textbf{b}_\textbf{k}=\textbf{a}_\textbf{k}^\dagger\mathcal{E}_\textbf{k}\textbf{a}_\textbf{k}.
\end{equation}
Hence we can identify the diagonalized matrix
\begin{equation}
    \mathcal{E}_\textbf{k}=P_\textbf{k}^\dagger \mathcal{H}_\textbf{k} P_\textbf{k}=\text{diag}(\epsilon_{\textbf{k},1},... \epsilon_{\textbf{k},N},\epsilon_{\textbf{-k},1},...,\epsilon_{\textbf{-k},N}),
\end{equation}
containing the normal boson mode energies
\begin{equation}
    H_{\rm AF}=\sum_k\sum_i \epsilon_{\textbf{k},i} (a^\dagger_{\textbf{k},i}a_{\textbf{k},i}+\frac{1}{2}).
    \end{equation}
Since we have the following identity on the transformation 
\begin{equation}\label{PGP=G}
    P_\textbf{k}\eta P_\textbf{k}^\dagger=\eta, 
\end{equation}
it is called \emph{para-unitary}, in analogy with a unitary transformation $U^\dagger I U=I$, and such a diagonalization is referred to as a \emph{para-unitary diagonalization}.

The algorithm that leads to $P_\mathbf{k}$ starts in \cite{Colpa1978} by assuming positive definiteness of the Hamiltonian and using the following theorem.
\begin{theorem}
A square hermitian matrix can be para-unitarily diagonalized into a matrix with all diagonal elements positive if and only if it is positive definite. 
\end{theorem}
This condition is required to perform a Cholesky decomposition on the coupling matrix,
\begin{equation}
\mathcal{H}_\textbf{k}=L L^\dagger ,
\end{equation}
where $L$ is a lower triangular matrix depending implicitly on momentum. We then diagonalize the matrix $W=L\eta L^\dagger$ with a unitary matrix $U$ such that we arrange the resulting diagonal matrix as such,
\begin{equation}
V=UWU^\dagger=\text{diag}(\epsilon_{\textbf{k},1},... \epsilon_{\textbf{k},N},-\epsilon_{\textbf{-k},1},...,-\epsilon_{\textbf{-k},N}).
\end{equation}
We hence find that $\mathcal{E}_\mathbf{k}=\eta V$ and verify straightforwardly that the matrix
\begin{equation}
P_\mathbf{k}=(L^\dagger)^{-1}U\mathcal{E}_\mathbf{k}^{1/2},
\end{equation}
fulfills the requirement $P^\dagger_\mathbf{k}\mathcal{H}_\mathbf{k} P_\mathbf{k}=\mathcal{E}_\mathbf{k}$. This concludes our problem for a positive definite coupling matrix. In the following sections, the $\mathbf{k}$-dependence of the BV matrix will be implicit. We will only reestablish the subscript in the final formula.

In many cases, such as the one treated in Section \ref{sec: application}, the band structure admits isolated points of 0-energy. These Goldstone modes can be treated by adding an infinitesimal positive number to the diagonal of the coupling matrix $\mathcal{H}_\mathbf{k}$. The latter solution is equivalent to introducing an anisotropy to the system that can be removed at the end of the diagonalization.

\subsection{Interfacial coupling}

If we assume a one-to-one correspondence between the metallic and the magnetic lattice sites at the interface between the materials, interfacial coupling can be modeled by
\begin{equation}
    H_{\rm int}=-\sum_{m,i\in \text{Int}} J_{i} \mathbf{s}_{m,i}.\mathbf{S}_{m,i},
\end{equation}
where it is sufficient to write $\mathbf{R}=\mathbf{R}_m+\mathbf{R}_i$ for both materials and $\text{Int}$ corresponds to the lattice sites at the interface. In momentum space, with $\mathbf{S}_{\mathbf{k},i}=\frac{1}{\sqrt{N}}\sum_{m}\mathbf{S}_{m,i}e^{i\mathbf{k} \mathbf{R}_{m}}$, it is written as
\begin{equation}
    H_{\rm int}=-\frac{1}{N}\sum_{i,\mathbf{q},\mathbf{k}} J_{\mathbf{q}\mathbf{k},i} \mathbf{s}_{\mathbf{q},i}.\mathbf{S}_{\mathbf{k},i}.
\end{equation}
This spin exchange Hamiltonian is responsible for driving the system in a non-equilibrium state. It will be expanded in a Keldysh evolution operator. 

\section{Dynamical spin susceptibilities}\label{sec: dynamical spin susceptibility}

\subsection{Metallic Green functions}

The retarded component of the spin susceptibility of the metallic part is defined as \cite{Kato2019}
 \begin{equation}\label{greeneleckato}
\begin{split}
  \chi^R_i(\mathbf{q},t)=&i\hbar^{-1}\Theta(t)\langle \left[ s_{\mathbf{q},i}^+ (t)  ,s_{\mathbf{q},i}^-(0)\right]  \rangle,\\=&\Theta(t)\left[\chi^>_i(\mathbf{q},t)-\chi^<_i(\mathbf{q},t)\right],
  \end{split}
\end{equation}
with $s^\pm=s^x\pm i s^y$. The lesser and greater components are defined
 \begin{equation}\label{chilesser}
\begin{split}
\chi^<_i(\mathbf{q},t)&=i\hbar^{-1}\langle s_{\mathbf{q},i}^- (0)s_{\mathbf{q},i}^+(t)  \rangle, \\ \chi^>_i(\mathbf{q},t)&=i\hbar^{-1}\langle s_{\mathbf{q},i}^+ (t)s_{\mathbf{q},i}^-(0)  \rangle .
  \end{split}
\end{equation}
In frequency domain we adopt
    \begin{equation}
\begin{split}
&\chi(\mathbf{q},t)=\int \frac{d\omega}{2\pi} e^{-i\omega t} \chi(\mathbf{q},\omega).
\end{split}
\end{equation}
Their relation to distribution function are given by
\begin{equation}\label{chidistib}
\begin{split}
     \chi^<_i(\mathbf{q},\omega)&=2if^{\rm NM}(\mathbf{q},\omega)\text{Im}\chi^R_i(\mathbf{q},\omega),\\
     \chi^>_i(\mathbf{q},\omega)&=2 i [1+f^{\rm NM}(\mathbf{q},\omega)]\text{Im}\chi^R_i(\mathbf{q},\omega),
\end{split}
\end{equation}
where $f^{\rm NM}$ is the (possibly non-equilibrium) distribution function that needs to be evaluated. The retarded Green function is explicitly given by \cite{Ohnuma2013, Adachi2013}
\begin{equation}
\chi^R(\textbf{q},\omega)=\frac{\chi_N}{1 + \lambda_N^2|\textbf{q}|^2-i\omega\tau_{sf}},
\end{equation}
with $\chi_N$, $\lambda_N$ and $\tau_{sf}$ being the
paramagnetic susceptibility, the spin diffusion length and the spin relaxation time in the metal. Note furthermore that this Green function is space independent and the $i$-index has been dropped. 

\subsection{Magnonic Green functions}\label{sec: green mag}
Similarly, we can define the retarded component of the spin susceptibility for the magnet
 \begin{equation}\label{greeneleckato}
\begin{split}
  \mathcal{G}^R_i(\mathbf{k},t)=&i\hbar^{-1}\Theta(t)\langle \left[ S_{\mathbf{k},i}^+ (t)  ,S_{\mathbf{k},i}^-(0)\right]  \rangle .
  \end{split}
\end{equation}
This is however the following magnon greater and lesser functions that will be used,
\begin{equation}
\begin{split}
G^<_{i}(\mathbf{k},t)&=-i\hbar^{-1}\langle a_{\mathbf{k},i}^\dagger(0)a_{\mathbf{k},i}(t) \rangle, \\
G^>_{i}(\mathbf{k},t)&= -i\hbar^{-1}\langle a_{\mathbf{k},i}(t)a^\dagger_{\mathbf{k},i}(0) \rangle ,
 \end{split}
\end{equation}
with $1\leqslant i \leqslant N$. Accordingly we make the necessary arrangement to express the two-point function
\begin{equation}\label{Glesser}
\begin{split}
    \langle (\mathbf{a}^\dagger_{\mathbf{k}})_j(0)  (\mathbf{a}_{\mathbf{k}})_j(t)\rangle=& i\hbar \left[G^<_{ [j]}(\eta\mathbf{k},t)\delta_{\eta,1}+ G^>_{ [j]}(\eta\mathbf{k},-t)\delta_{\eta,-1}\right],\\
    =&i\hbar G^{\eta<}_{[j]}(\eta\mathbf{k},\eta t),
\end{split}
\end{equation}
where $1\leqslant j \leqslant 2N$. The notation $[j]=(j-1 \mod N) +1$ takes care of the doubling of operators. Furthermore we have written $\eta$ instead of $\eta_{j}\equiv\eta_{jj}$ and $\eta$ is as defined in Sec.~\ref{Colpa}. When $\eta_{j}=-1$ the notation $G^{\eta_{j}<}$ and $G^{\eta_{j}A}$ refers respectively to $G^>$ and $G^R$. Equivalently,
\begin{equation}\label{G--}
\begin{split}
    \langle \Tilde{T} (\mathbf{a}^\dagger_{\mathbf{k}})_j(0)  (\mathbf{a}_{\mathbf{k}})_j(t)\rangle= & \langle \Tilde{T} a_{\mathbf{k},j}^\dagger(0)a_{\mathbf{k},j}(t) \rangle \delta_{\eta,1} \\&+ \langle \Tilde{T} a_{-\mathbf{k},[j]}(0)a^\dagger_{-\mathbf{k},[j]}(t) \rangle\delta_{\eta,-1},\\=& i\hbar G^{--}_{[j]}(\eta\mathbf{k},\eta t),
\end{split}
\end{equation}
where $\Tilde{T}$ denote anti time ordering. The retarded component is then obtained from the identity
\begin{equation}\label{Gredef}
\begin{split}
G^{--}=G^{\eta_j<}-G^{\eta_j A}.
\end{split}
\end{equation}
Following our convention we define the Fourier transform
\begin{equation}
\begin{split}
 G(\mathbf{k},t)=\int \frac{d\omega}{2\pi} e^{-i\omega t} G(\mathbf{k},\omega),
\end{split}
\end{equation}
and the relation
\begin{equation}\label{Gdistib}
\begin{split}
     G^{\eta_j<}_{[j]}(\mathbf{k},\omega)&=2 i [ f^{\rm MI}(\mathbf{k},\omega)+\frac{1}{2}(1-\eta_j)]\text{Im} G^{R}_{[j]} (\mathbf{k},\omega),\\
\end{split}
\end{equation}
summarizes the role of $\eta_{j}$. An explicit formula for the equilibrium retarded Green function is derived as ($\hbar=1$)
\begin{equation}\label{eq:green function of magnons}
G_i^R(\textbf{k},\omega)=\frac{2 S_0}{\omega-\epsilon_{\mathbf{k},i}+i\vartheta},
\end{equation}
with $\vartheta$ the broadening.

Here $f^{\rm MI}$ needs also be replaced by equilibrium or non equilibrium distribution function. In the case of the  spin-Seebeck effect, both $f^{\rm MI}$ and $f^{\rm NM}$ can be replaced by the Bose-Einstein distribution $n_B(\omega,T)=(e^{\frac{\omega}{k_B T}}-1)^{-1}$ or the Wigner distribution \cite{Kamenev2009}. For other non-equilibrium dynamics, such as spin-pumping with FMR-drive, a non-equilibrium distribution $f^{\rm MI}$ must be evaluated in perturbation with respect to $V$.

\section{Interfacial spin current}\label{sec: spin-current}
\subsection{Full generality formulation}
We wish to evaluate the spin current at the NM-MI interface for the general system described in the previous section. It is expressed through the Heisenberg equation
\begin{equation}
\hat{\mathbf{I}}=\sum_{{\bf R} \in \text{Int}}i [ \mathbf{s}({\bf R}), H_{\rm int}],
\end{equation}
the vector index of $\hat{\mathbf{I}}$ ($\alpha=\{x,y,z\}$) indicating the polarization. Focusing on one of these components
\begin{equation}
\begin{split}
    \hat{I}^\alpha &=\sum_{{\bf R} \in \text{Int}, \beta, \gamma} 2J_{{\bf R}} \epsilon_{\alpha \beta \gamma} s^\gamma({\bf R}) S^\beta({\bf R}).\\
        \end{split}
\end{equation}
From now on ${\bf R}=\textbf{R}_i\in \text{Int}$ is always assumed and we omit it (note that the same symbol $i$ is used for both a subscript for lattice site and the imaginary number). In momentum space we obtain
\begin{equation}
\begin{split}
    \hat{I}^\alpha 
    &=\sum_{  \mathbf{q} \mathbf{k},i,\beta \gamma} 2J_{\mathbf{q},\mathbf{k},i}\epsilon_{\alpha \beta \gamma} s_{\mathbf{q},i}^\gamma   S_{\mathbf{k},i}^\beta .
    \end{split}
\end{equation}
We verify easily that it yields the $(\alpha=z)$-polarized result of a Heisenberg ferromagnet $H_{\rm FI}=\sum_{\langle i,j\rangle}J_{ij} \mathbf{S}_{i}. \mathbf{S}_{j}$, that is, $ I^z =-i\sum_{ \mathbf{q} \mathbf{k}} J_{\mathbf{q},\mathbf{k}} 
     S_{\mathbf{k}}^+
     s_{\mathbf{q}}^-   -\text{H.c}$. Taking the average and dropping the position dependence of the exchange coupling,
\begin{equation}
\begin{split}
    \langle\hat{I}^\alpha  \rangle& = \lim\limits_{t_1,t_2 \rightarrow 0} \sum_{  \mathbf{q} \mathbf{k},i,\beta \gamma} 2J_{\mathbf{q},\mathbf{k}}\epsilon_{\alpha \beta \gamma} \langle s_{\mathbf{q},i}^\gamma(t_2)   S_{\mathbf{k},i}^\beta(t_1) \rangle.\\
    \end{split}
\end{equation}

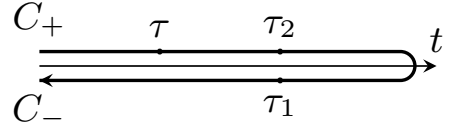
\begin{figure}[t]
\centering
\scalebox{1.6}{
    \begin{tikzpicture}[scale=1]
    
    \path[draw,line width=0.8pt]
    (-3,0.12) node[above] {$C_+$} -- 
    (-2,0.12) node[above] {$\tau$} --
    (-1,0.12) node[above] {$\tau_2$} --
    (0,0.12) 
    arc (90:-90:0.12) --
    (0,-0.12)--
    (-1,-0.12) node[below] {$\tau_1$} --
    (-3,-0.12) node[below] {$C_-$} ;
    \filldraw [black] (-2,0.12) circle (0.5pt);
        \filldraw [black] (-1,0.12) circle (0.5pt);
    \filldraw [black] (-1,-0.12) circle (0.5pt);
    \draw [-{stealth[scale=20]}](-2.9,-0.12) -- (-3,-0.12);
    \draw [-stealth] (-3,0) -- (.3,0) node[above] { $t$}; 
    \end{tikzpicture}}
\caption{The Keldysh time contour $c$ consist in two branches $c=C_+ \cup C_-$. All time points of the upper branch $C_+$ are earlier than times of the lower branch $C_-$. }
\label{fig:kato}
\end{figure}
We use Keldysh perturbation expansion \cite{Rammer1986} and replace the limit by setting $\tau_{1(2)}$ on the lower $C_-$ (upper $C_+$) branch of the Keldysh contour, see Fig. \ref{fig:kato}. The $\alpha$-polarized spin current through the interface reads
    \begin{equation}
\begin{split}
    \langle\hat{I}^\alpha \rangle= \sum_{  \mathbf{q} \mathbf{k},i,\beta \gamma} &\epsilon_{\alpha \beta \gamma}\left\langle  T_c \left[  s_{\mathbf{q},i}^\gamma(\tau_2)   S_{\mathbf{k},i}^\beta(\tau_1) \right.\right. \\&\times \left. \left. \exp\left(-\frac{i}{\hbar}\int_c d\tau H_{int}(\tau)\right)\right] \right\rangle_0,
\end{split}
\end{equation}
where the average is taken over the unperturbed system. We expand Keldysh evolution operator to lowest order. We end up with $\langle I^\alpha \rangle$ to second order in $J$,
\begin{equation}\label{generalcurr}
\begin{split}
\langle\hat{I}^\alpha  \rangle=&\int_c d\tau \sum_{\mathbf{q} \mathbf{q}'}\sum_{\mathbf{k} \mathbf{k}'}\sum_{i,j}\sum_{\beta \gamma \delta}  \frac{2}{\hbar}J_{\mathbf{q},\mathbf{k}}J_{\mathbf{q}',\mathbf{k}'}\epsilon_{\alpha \beta \gamma} \\&\times \left\langle T_c  s_{\mathbf{q},i}^\gamma (\tau_2)  s_{\mathbf{q}',j}^\delta(\tau)
 \right\rangle_0 
 \left\langle T_c  S_{\mathbf{k},i}^\beta (\tau_1)  S_{\mathbf{k}',j}^\delta(\tau)
 \right\rangle_0.
\end{split}
\end{equation}
This formula has been obtained in full generality. In the context of a simple realistic system (non magnetic metal), we prove in Appendix~\ref{elec} and \ref{sec: Current contributions}  that we can express the correlators in terms of $s^\pm$ and bring some restriction on the $\gamma$ index of the insulator's spin operators. We hence reduce it to
\begin{equation}\label{current intermediaire}
\begin{split}
    \langle \hat{I}^\alpha \rangle=&\text{Re}\Bigg[- \frac{i}{\hbar}\sum_{\mathbf{q}\mathbf{k},i, \beta, \gamma=\{x,y\}} |J_{\mathbf{q}\mathbf{k}}|^2\epsilon_{\alpha \beta \gamma} \\& \int_c d\tau\langle T_c s^-_{\mathbf{q},i}(\tau_2)s^+_{\mathbf{q},i}(\tau)\rangle_0\langle  T_c S^\beta_{\mathbf{k},i}(\tau_1)S^\gamma_{\mathbf{k},i}(\tau)\rangle_0 \Bigg].
 \end{split}
\end{equation}
with $\beta$ taking value in $\{x,y,z\}$. It is then expressed in the real-time representation \cite{Rammer1986} (with $t_1=t_2=0$) as
\begin{equation}\label{currentrealtime}
\begin{split}
    \langle\hat{I}^\alpha \rangle
    =\text{Re}\Bigg[ & - \frac{i}{\hbar}\sum_{\mathbf{q}\mathbf{k},i, \beta, \gamma=\{x,y\}} |J_{\mathbf{q}\mathbf{k}}|^2\epsilon_{\alpha \beta \gamma} \\ \times& \int dt \left[ \langle T s^-_{\mathbf{q},i}(0)s^+_{\mathbf{q},i}(t)\rangle_0\langle S^\beta_{\mathbf{k},i}(0)S^\gamma_{\mathbf{k},i}(t)\rangle_0 \right.  \\&  - \left.\langle s^+_{\mathbf{q},i}(t)s^-_{\mathbf{q},i}(0)\rangle_0\langle \Tilde{T} S^\beta_{\mathbf{k},i}(0)S^\gamma_{\mathbf{k},i}(t)\rangle_0 \right] \Bigg].
    \end{split}
\end{equation}
In Appendix ~\ref{MagnonApp}, we explain how to write the quantity $\sum_{ \beta, \gamma=\{x,y\} }\epsilon_{\alpha \beta \gamma}\left\langle (\Tilde{T}) S_{\mathbf{k}}^\beta (t_1)  S_{\mathbf{k}}^\gamma(t_2)
 \right\rangle_0$ in terms of Eqs. (\ref{Glesser}) and  (\ref{G--}). The formulation of Eq. (\ref{currentrealtime}) in terms of the previously defined Green functions of the electronic and magnetic systems, Eqs.~(\ref{chilesser}), (\ref{Glesser}), (\ref{G--}), follows straightforwardly
\begin{equation}
\begin{split}
\langle\hat{I}^\alpha\rangle=\text{Re}&\bigg [  \frac{\hbar }{4}\sum_{\mathbf{q}\mathbf{k},ij} |J_{\mathbf{q},\mathbf{k}}|^2 (P^\dagger_\mathbf{k}\Upsilon^{\alpha,i} P_\mathbf{k})_{jj}\\&\times\int dt \left[ \chi^{++}_i(\mathbf{q},t) G^{\eta_j<}_{[j]}(\eta_j\mathbf{k},\eta_j t)\right.\\& \left. \qquad\quad - \chi^>_i(\mathbf{q},t) G^{--}_{[j]}(\eta_j\mathbf{k},\eta_j t) \right] \bigg ],
\end{split}
\end{equation}
with $\Upsilon^{\alpha,i}$ a diagonal matrix whose only non-zero elements are
\begin{equation}
\begin{split}
\Upsilon^{\alpha,i}_{i,i}=&\sum_{ \beta, \gamma=\{x,y\}}2 \epsilon_{ \alpha \beta\gamma} S_i  \Im u_i^\beta u^{*\gamma}_i,\\
\Upsilon^{\alpha,i}_{N+i,N+i}=&\sum_{\beta, \gamma=\{x,y\}}2 \epsilon_{ \alpha \beta \gamma} S_i   \Im  u_i^{*\beta} u^\gamma_i=-\Upsilon^{\alpha,i}_{i,i}.
\end{split}
\end{equation}
We then use $\chi^{++}=\chi^A+\chi^>$ and Eq.~(\ref{Gredef}), and obtain in Fourier space
\begin{equation}
\begin{split}
    \\\langle\hat{I}^\alpha\rangle=\text{Re}&\bigg[\frac{\hbar}{4 }\sum_{\mathbf{q}\mathbf{k},ij, \beta, \gamma=\{x,y\}} |J_{\mathbf{q},\mathbf{k}}|^2(P^\dagger_\mathbf{k}\Upsilon^{\alpha,i} P_\mathbf{k})_{jj}\\&\times\int \frac{d\omega}{2\pi} \left[\chi^A_i(\mathbf{q},\omega) G^{\eta_j<}_{[j]}(\eta_j\mathbf{k},-\eta_j\omega)\right.\\&\left.\qquad\qquad + \chi^>_i(\mathbf{q},\omega) G^{\eta_j A}_{[j]}(\eta_j\mathbf{k},-\eta_j\omega) \right] \bigg ].
\end{split}
\end{equation}
Using Eqs.~(\ref{chidistib}) and (\ref{Gdistib}) we finally arrive at
\begin{equation}\label{Current most general}
\begin{split}
    \langle\hat{I}^\alpha\rangle=&\frac{\hbar}{2}\sum_{\mathbf{q}\mathbf{k},ij} |J_{\mathbf{q}\mathbf{k}}|^2(P^\dagger_\mathbf{k}\Upsilon^{\alpha,i} P_\mathbf{k})_{jj} \\ &\times \int \frac{d\omega}{2\pi} \text{Im}\chi^R_i(\mathbf{q},\omega) \text{Im} G^{R}_{[j]}(\eta_j\mathbf{k},-\eta_j\omega)\\&\times\left[f^{\rm MI}(\eta_j \mathbf{k},-\eta_j\omega) + \eta_j f^{\rm NM}(\mathbf{q},\omega) +\frac{1}{2}(1+\eta_j) \right].
\end{split}
\end{equation}
This formula is the central result of this Article and expresses the spin current pumped by interfacial magnons assuming the most general magnonic Hamiltonian. The distribution functions of the magnetic insulator and the normal metal should then be replaced according to the setup under consideration.

\subsection{Spin Seebeck effect}
For thermal spin-injection, we can replace the distribution function by the Bose-Einstein distribution.
We recall $\eta_j=1$ for $j\leq N$ and $-1$ for $j> N$. Using the equality $n_B(-\eta_j\omega,T) +\frac{1}{2}(1+\eta_j)=-\eta_j n_B(\omega,T)$, and a parity evaluation with respect to $\omega$, we obtain
\begin{equation}\label{current final}
\begin{split}
    \langle\hat{I}^\alpha\rangle=&\frac{\hbar}{2}\sum_{\mathbf{q}\mathbf{k},ij} |J_{\mathbf{q},\mathbf{k}}|^2(P^\dagger_\mathbf{k}\Upsilon^{\alpha,i} P_\mathbf{k})_{jj}
     \\&\times\int \frac{d\omega}{2\pi} \text{Im}\chi^R_i(\mathbf{q},\omega) \text{Im} G^{R}_{[j]}(\eta_j\mathbf{k},\omega)\\&\times\eta_j\left[ n_B(\omega,T_{\rm NM})-n_B(\omega,T_{\rm MI})\right],
\end{split}
\end{equation}
where $T_{\rm MI},\;T_{\rm NM}$ are the temperatures of the magnetic insulator and the normal metal, respectively. This result is consistent with the fact that spin injection does not occur when $T_{\rm MI}=T_{\rm NM}$. Expression (\ref{current final}) is a generalization of the one obtained for collinear magnets \cite{Adachi2013,Kamra2017,Matsuo2018,Kato2019,Ohnuma2013}. The difference is the presence of weights depending on the ground state configuration of the insulating magnet, and that can be gathered in a new matrix with elements
\begin{equation}
\begin{split}
    \mathsf{M}^\alpha_{i,j}({\mathbf{k}})=(\eta P_{\mathbf{k}}^\dagger \Upsilon^{\alpha,i}P_{\mathbf{k}})_{jj}.
\end{split}
\end{equation}
Note that $ \mathsf{M}^\alpha(\mathbf{k})$ is not square since $i$ runs over the lattice sites at the interface while $j$ runs over the energy eigenstates of the magnonic system. This term is the only one bearing the polarization index of the current and geometrical information on spin configuration of the insulator's ground state.

In analogy with the Landauer-Büttiker formalism \cite{Landauer1957,Bttiker1985} we identify the transmission coefficient for $\alpha$-polarized spin-current. Replacing $J_{\mathbf{q},\mathbf{k}}=J$ for simplicity, it is expressed
\begin{equation}\label{eq: transmission coeff}
\begin{split}
    \mathcal{T}^\alpha(\omega)=\frac{|J|^2}{2}
     \sum_{\mathbf{q},\mathbf{k}}\sum_{i,j}&  \mathsf{M}^\alpha_{i,j}({\mathbf{k}}) \text{Im}\chi^R_i(\mathbf{q},\omega) \\&\times\text{Im} G^{R}_{[j]}(\eta_j\mathbf{k},\omega),
\end{split}
\end{equation}
from which we write the current in a proper Landauer-like form \cite{Caroli1971},
\begin{equation}\label{landauercurr1}
\begin{split}
\langle\hat{I}^\alpha\rangle=\hbar\int \frac{d\omega}{2\pi}\mathcal{T}^\alpha(\omega)\left[n_B(\omega,T_{\rm MI}) -n_B(\omega,T_{\rm NM}) \right].
\end{split}
\end{equation}

This result covers thermal spin-injection to second order in the interfacial coupling under assumption of linear spin wave theory, for all magnets and for metals treated as spin-sinks without spin-orbit coupling.

We now show for consistency that, starting from Eq. (\ref{landauercurr1}), we can derive the usual ferromagnetic case. Indeed, using Eq.~\eqref{PGP=G}, $\chi_i=\chi$ and the parity of $G^R$ in momentum ($\mathcal{E}_\mathbf{k}=\mathcal{E}_\mathbf{-k}$), we get $\sum_i  \mathsf{M}_{ij}^z=2\eta_j^2=2$ and $\mathcal{T}^z(\omega)=\sum_{\mathbf{q k}j}\text{Im}\chi^R(\mathbf{q},\omega)\text{Im} G^{R}_{[j]}(\mathbf{k},\omega)$, which finally gives
\begin{equation}\label{eq:ferro_diagonal_in_space}
\begin{split}
\langle\hat{I}^z\rangle=A\int \frac{d\omega}{2\pi}\sum_{\mathbf{q},\mathbf{k}}&\sum_{j} \text{Im}\chi^R(\mathbf{q},\omega)\text{Im} G^{R}_{[j]}(\mathbf{k},\omega)\\& \times \left[n_B(\omega,T_{\rm MI}) -n_B(\omega,T_{\rm NM}) \right],
\end{split}
\end{equation}
with $A=N\hbar|J|^2/2$.

\subsection{Spin pumping}

Pumping with a FMR drive is modeled by adding $V\neq 0$ in the Hamiltonian. This term is treated perturbatively and a correction to the lesser Green function
\begin{equation}\label{landauercurr2}
\begin{split}
\delta G^< (\mathbf{k},\omega)=G^R (\mathbf{k},\omega)\Sigma(\mathbf{k},\omega)G^A (\mathbf{k},\omega),
\end{split}
\end{equation}
yields a new distribution $f^{\rm MI}( \mathbf{k},\omega)=n_B(\omega,T_{\rm MI})+\delta f^{\rm MI} (\mathbf{k},\omega)$, with

\begin{equation}\label{delta fMI}
\begin{split}
\delta f^{\rm MI} (\mathbf{k},\omega)=\delta G^< (\mathbf{k},\omega)/[2i \text{Im}G^R (\mathbf{k},\omega)].
\end{split}
\end{equation}
Starting from Eq.~(\ref{Current most general}) and setting both materials at equal temperature, the difference of Bose-Einstein distributions vanishes and we obtain,
\begin{equation}\label{current_pump}
\begin{split}
   \langle\hat{I}^\alpha\rangle=\frac{|J|^2}{2}
     \sum_{\mathbf{q},\mathbf{k}}\sum_{i,j}&  \mathsf{M}^\alpha_{i,j}({\mathbf{k}})\int \frac{d\omega}{2\pi} \text{Im}\chi^R_i(\mathbf{q},\omega) \\\times\text{Im} G^{R}_{[j]}&(\eta_j\mathbf{k},-\eta_j\omega)\delta f^{\rm MI}_{[j]} (\eta_j\mathbf{k},-\eta_j\omega).
\end{split}
\end{equation}
The spin-current therefore inherits a Landauer form as in the spin Seebeck effect case.

\section{Application: the honeycomb topological magnon insulator}\label{sec: application}

\subsection{Model}

\begin{figure}[t]
\includegraphics[width=0.5\textwidth]{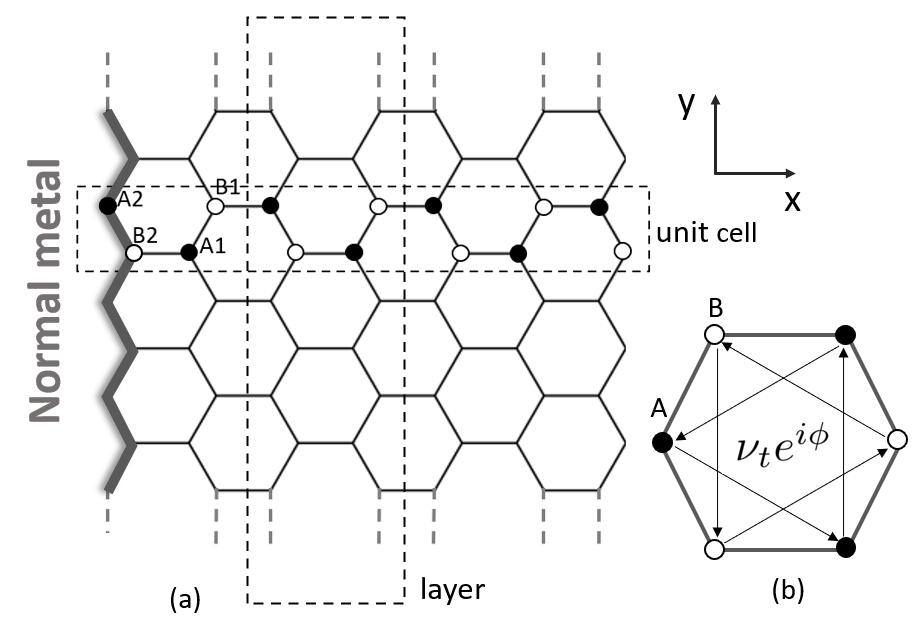}
\caption{ \label{fig:diaghopp} (a) Zig-zag edge ribbon infinite along $\hat{y}$. (b) Phase accumulation diagram.}

\end{figure}

For illustrative purpose we turn to the application of our formula Eq. (\ref{landauercurr1}) to a topological collinear ferromagnet. The following Hamiltonian, a bosonic analog of the Haldane model introduced in \cite{Owerre2016}, models a Honeycomb Heisenberg ferromagnet with second nearest neighbor interaction. In addition to a Heisenberg coupling, the lattice geometry naturally induces a Dzyaloshinskii-Moriya interaction
\begin{equation}\label{eq:spin hamilto owerre}
\begin{split}
H_{\rm MI}=-J\sum_{<lm>}&\mathbf{S}_l.\mathbf{S}_m-J'\sum_{\ll lm \gg}\mathbf{S}_l.\mathbf{S}_m\\&+\sum_{\ll lm \gg}\mathbf{D}_{lm}.(\mathbf{S}_l\times\mathbf{S}_m),
\end{split}
\end{equation}
with $\mathbf{D}_{lm}\propto \mathbf{R}_l\times \mathbf{R}_m$. Note that in Eq. (\ref{eq: transmission coeff}) the NM-MI coupling $J$ appears only in an overall factor. We formally set it equal to the Heisenberg coupling of Eq. (\ref{eq:spin hamilto owerre}), playing the role of an energy scale. In the linear spin-wave regime the Holstein-Primakoff transformation yields the bosonic description
\begin{equation}
\begin{split}
H_{\rm MI}=&\nu_0\sum_{l}b^\dagger_lb_l-\nu_s\sum_{<lm>}(b_l^\dagger b_m + h.c)\\&-
\nu_t\sum_{\ll lm\gg}[e^{i\phi_{lm}}b_l^\dagger b_m + h.c],
\end{split}
\end{equation}
with $\nu_0=3JS+6J'S$ (each ion has 3 nearest and 6 next nearest neighbors), $\nu_s=JS$, $S(J'+iD\nu_{lm})=\nu_t e^{i\phi_{lm}}$ and $\nu_{lm}=\pm1$ depending on the orientation with respect to the position of the third ion involved in the antisymmetric exchange. This yields the phase accumulation diagram of Fig. \ref{fig:diaghopp}, where the direction of the arrow indicates a positive phase multiplication. The topological properties are identified to be related to the complex parameter $\mathsf{Z}=S(J'+iD\nu_{lm})=\nu_t e^{i\phi_{lm}}$. The magnet enters its topological phase whenever $D>0$. See \cite{Owerre2016} for details.

\begin{figure*}

    \subfloat[$\mathbf{J'/J=0, D/J=0}$]{\includegraphics[width=0.32\textwidth]{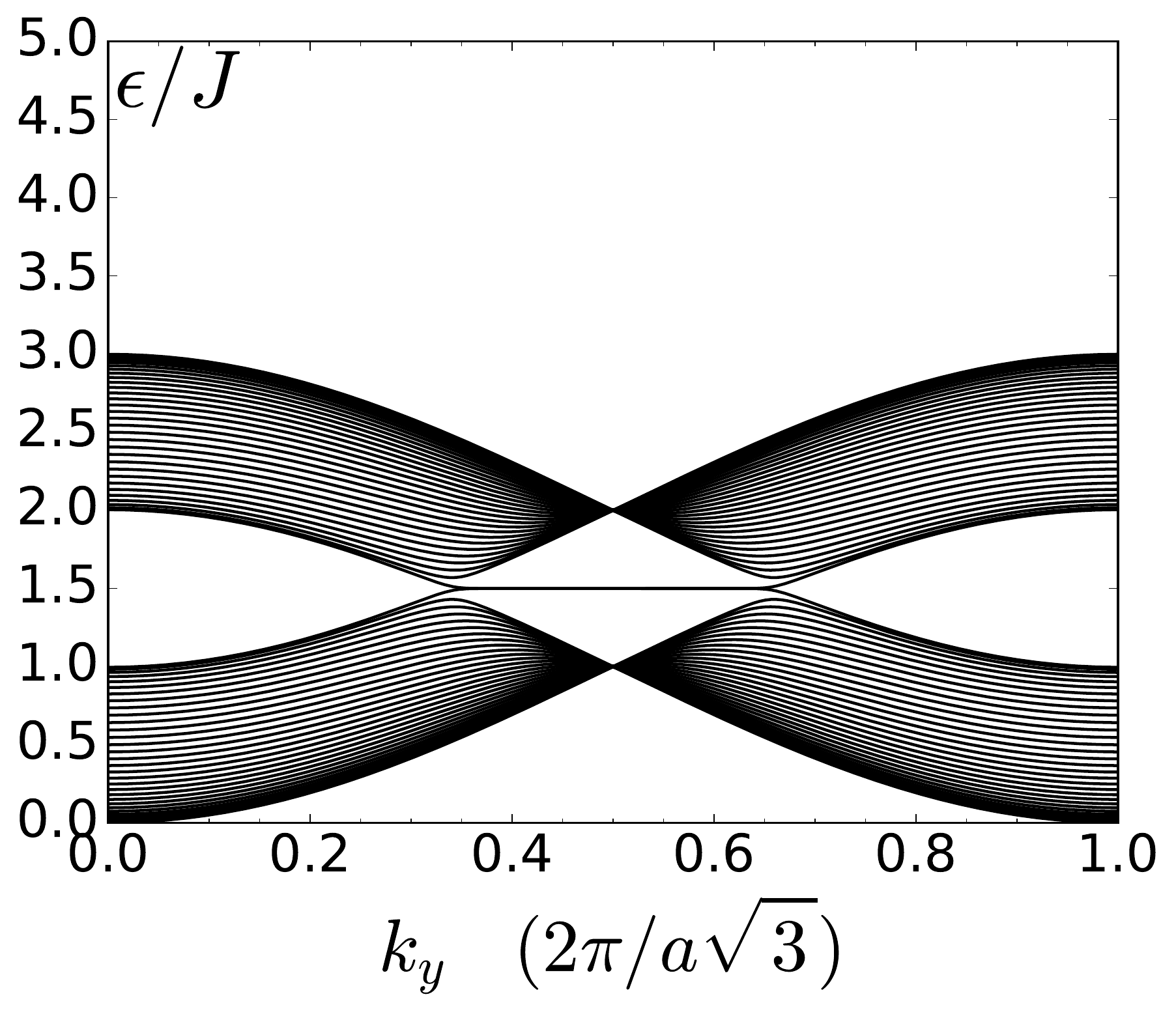} }%
\hfill
        \subfloat[$\mathbf{J'/J=0.6, D/J=0.2}$]{\includegraphics[width=0.32\textwidth]{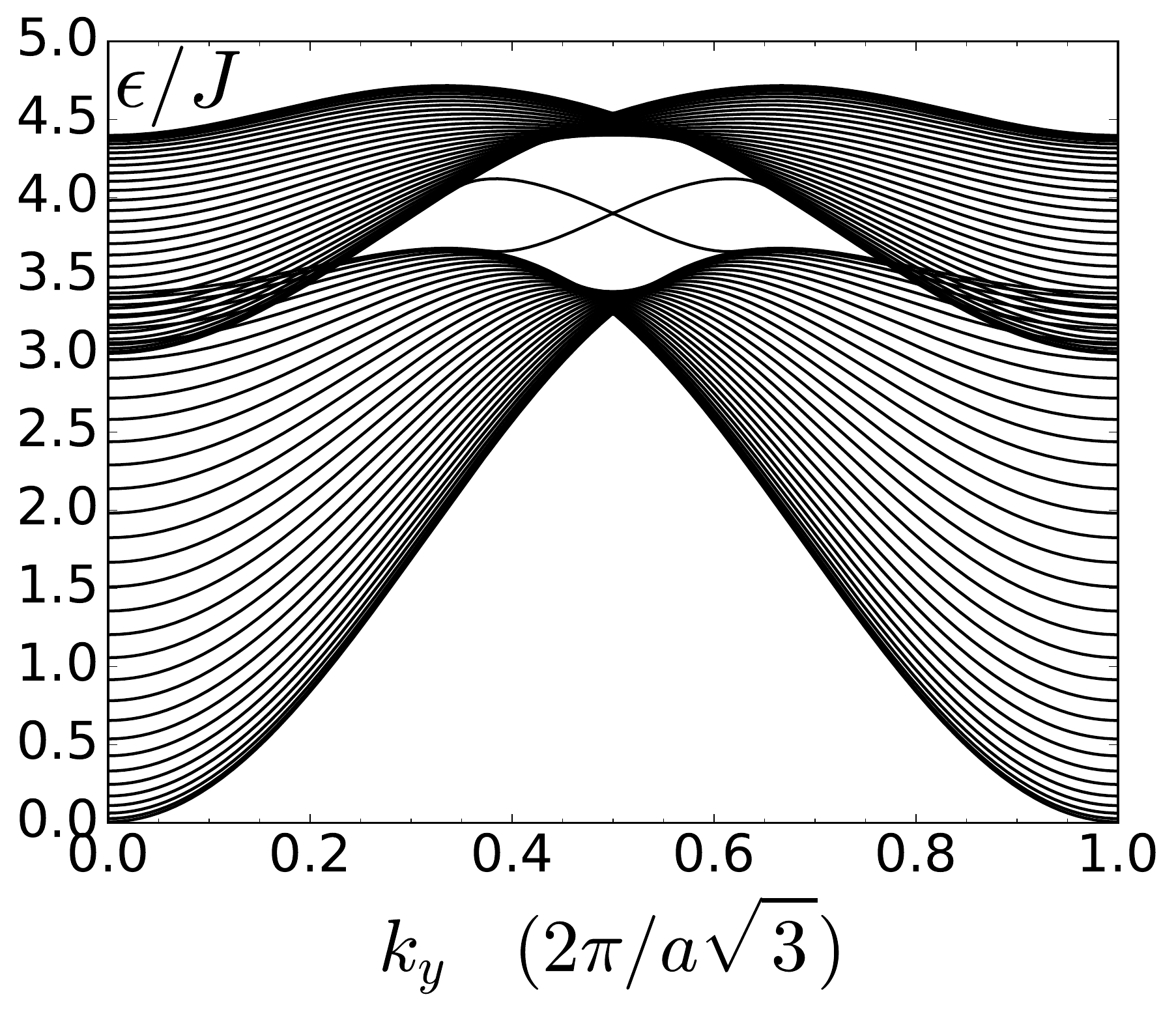} }%
\hfill
    \subfloat[$\mathbf{J'/J=0.23, D/J=0.6}$]{\includegraphics[width=0.32\textwidth]{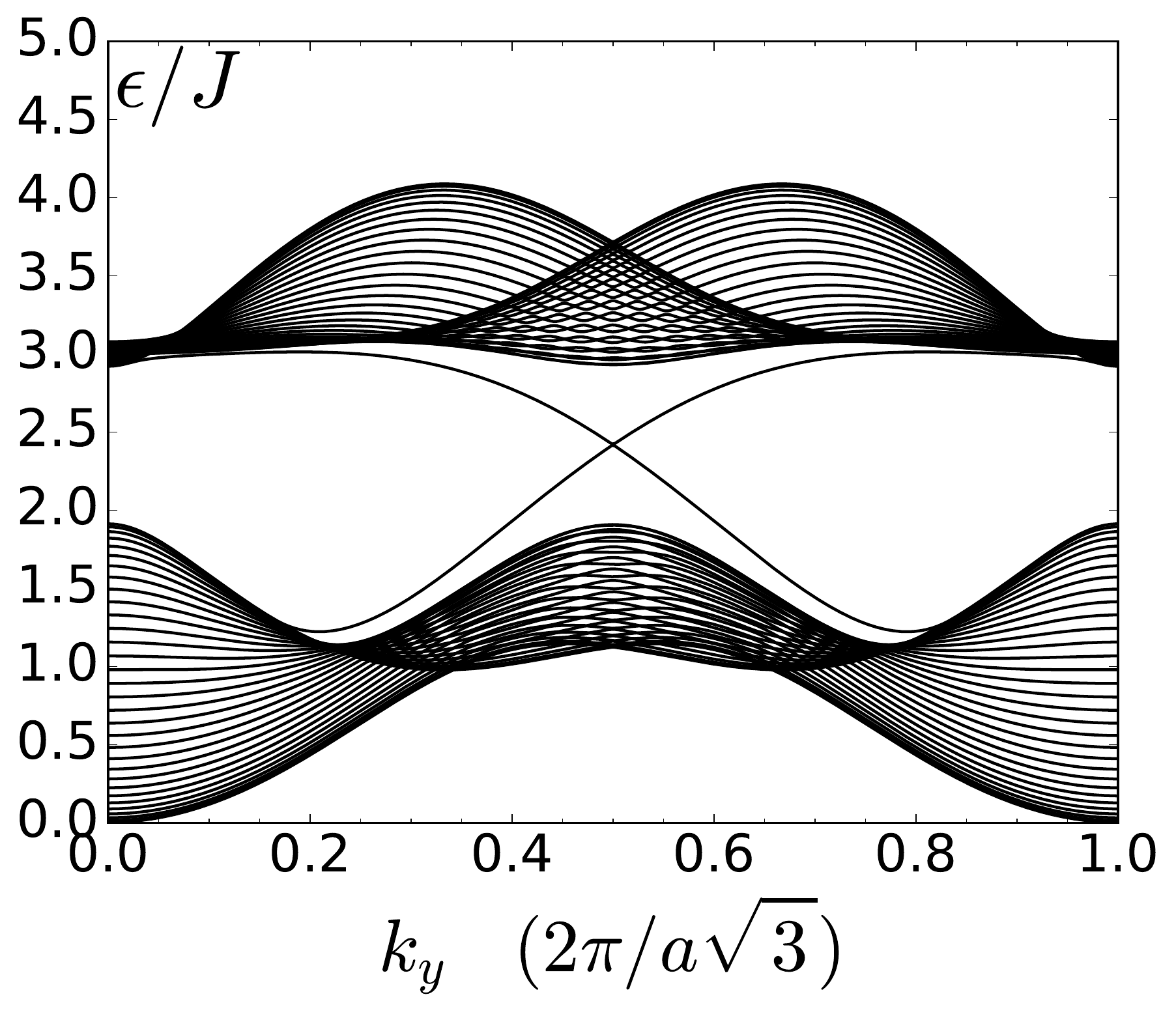} }%
\flushleft
    \subfloat[$\mathbf{J'/J=0, D/J=0}$]{{\includegraphics[width=0.32\textwidth]{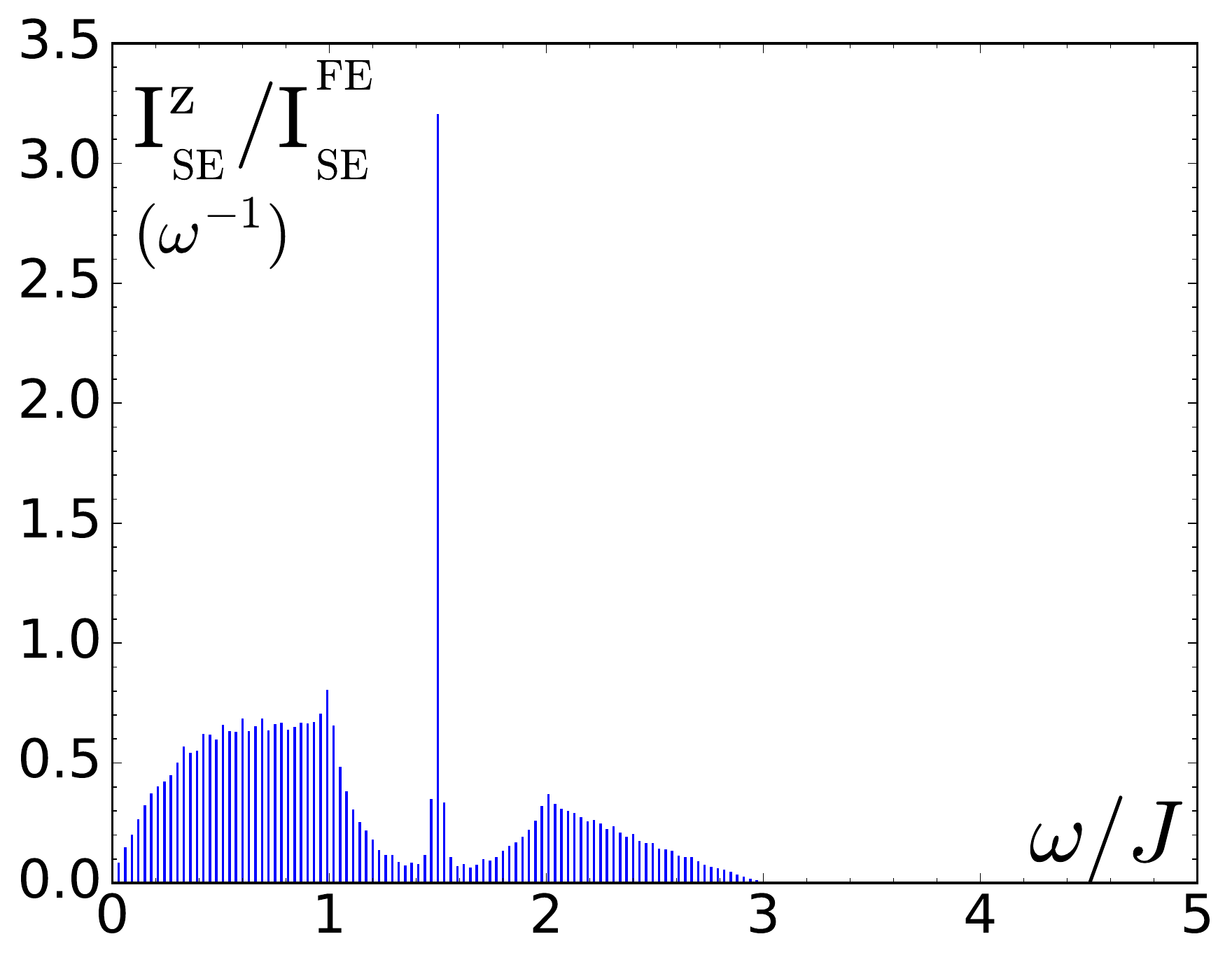} }}%
    \hfill
    \subfloat[$\mathbf{J'/J=0.6, D/J=0.2}$]{{\includegraphics[width=0.32\textwidth]{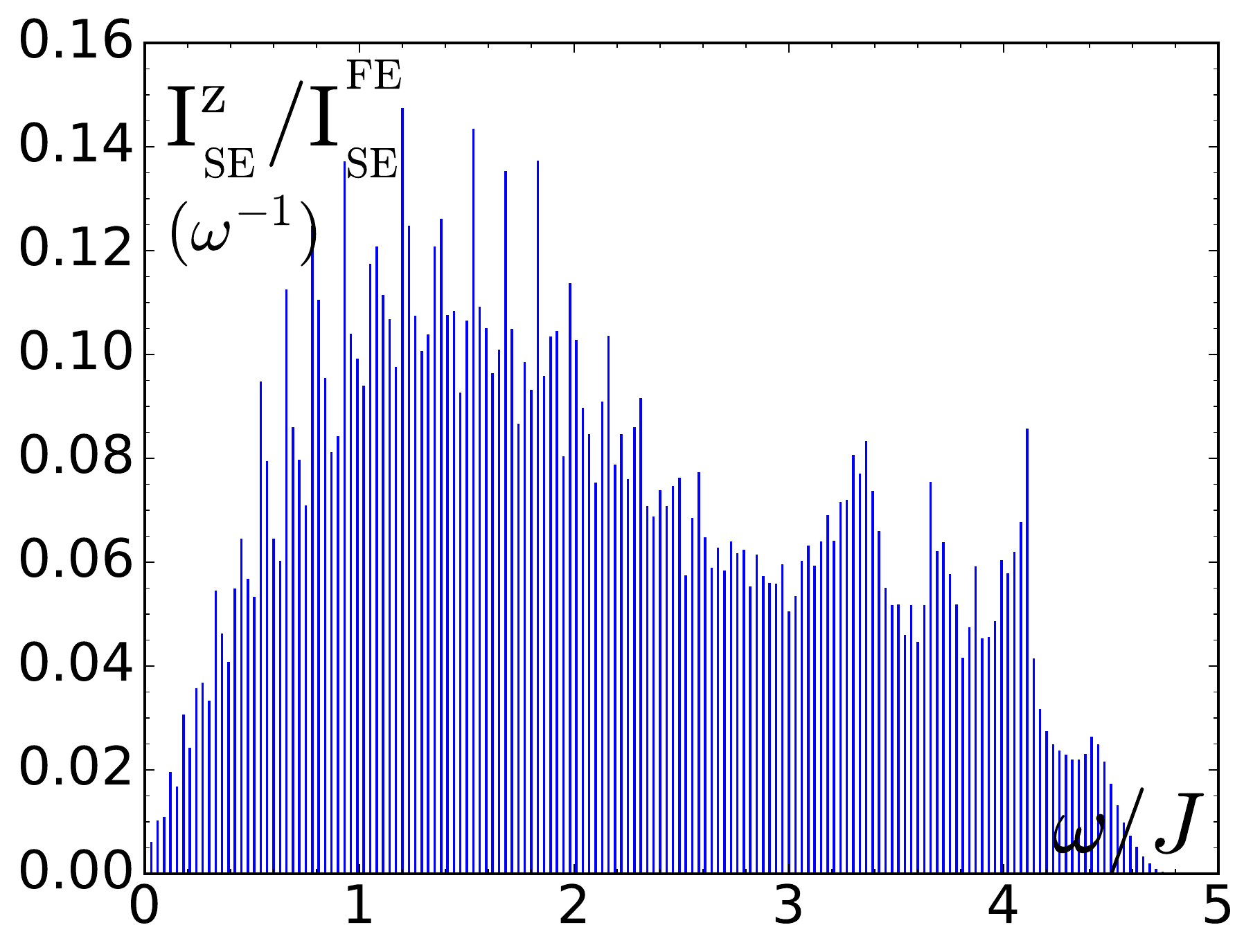} }}%
\hfill
    \subfloat[$\mathbf{J'/J=0.23, D/J=0.6}$]{{\includegraphics[width=0.32\textwidth]{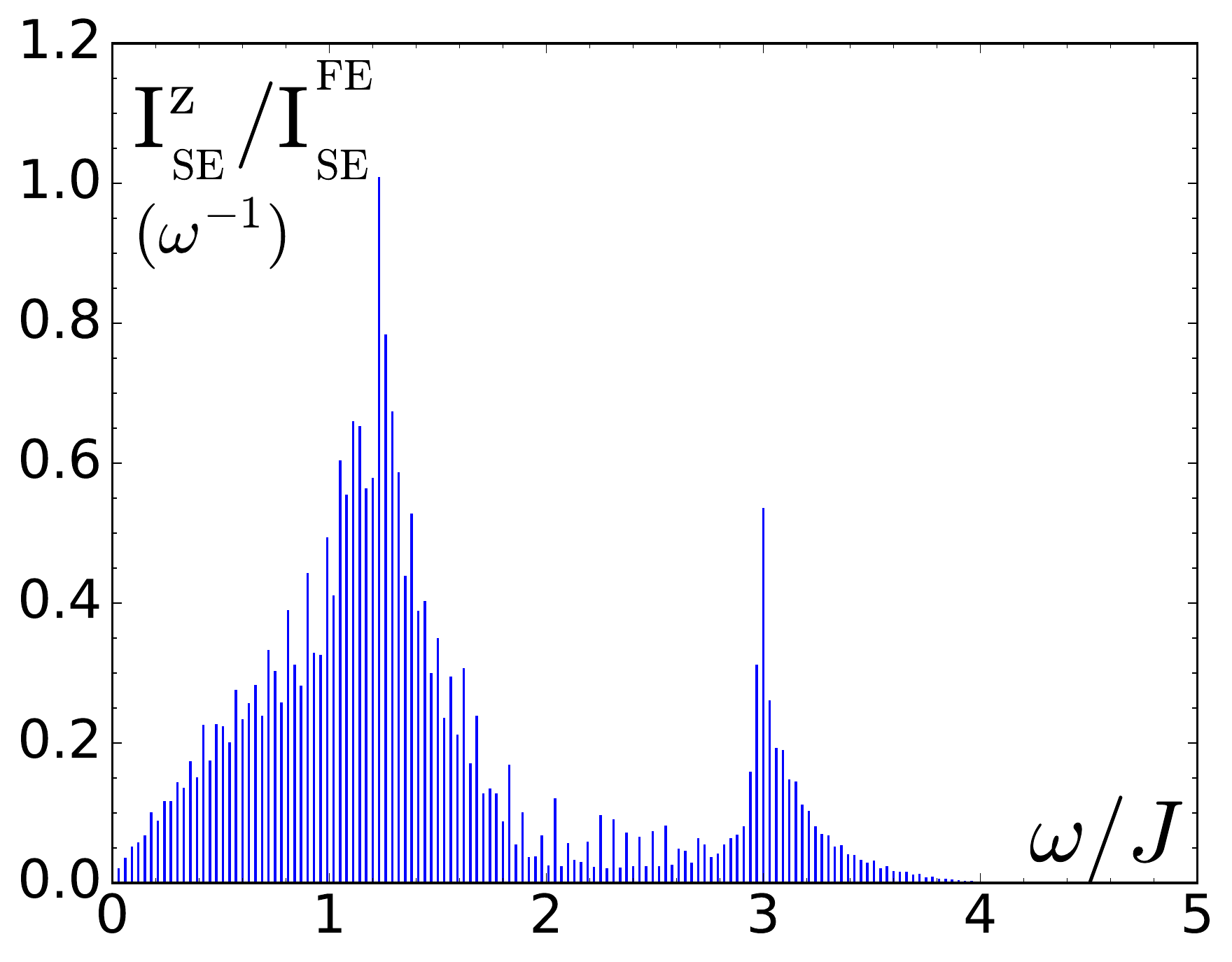} }}%
    
    \caption[]{Upper panels: energy bands for a  15 unit cells zig-zag nanoribbon (along $y$) of the topological honeycomb ferromagnet, for $J'$ and $D$ indicated bellow. Lower panels: frequency contribution to the spin-current generated by the spin-Seebeck effect; value normalized by the spin-current total of the ferromagnetic case $I^{\tiny FE}_{\tiny SE}=I^z_{\tiny SE}(J'=D=0)$. The ribbon is composed of 50 unit cells.}
    \label{fig:bands and seebeck}%
\end{figure*}

We study the interfacial spin pumping by building a nanoribbon along the $y$-axis (zig-zag edge). In order to use our formulation, we must express the Hamiltonian in the BdG form,
\begin{equation}\label{eq: BDG_Owerre}
\begin{split}
H_{\rm MI}=\frac{1}{2}\sum_{k_y}\Psi^\dagger_{k_y}\begin{pmatrix}\mathcal{J}_{k_y} & 0\\0&\mathcal{J}^*_{-k_y}\end{pmatrix}\Psi_{k_y},
\end{split}
\end{equation}
with $\Psi_{k_y}=\begin{pmatrix}\beta_{1,k_y}&\ldots&\beta_{n,k_y}&\beta_{1,-k_y}^\dagger&\ldots&\beta_{n,-k_y}^\dagger\end{pmatrix}^T$
and $\beta_{i,k_y}=\begin{pmatrix}a_{i,1,k_y}&a_{i,2,k_y}&b_{i,1,k_y}&b_{i,2,k_y}\end{pmatrix}^T$, since a unit cell admits four sub-lattices. The derivation of the nanoribbon Hamiltonian in Appendix \ref{Appendix: Owerre Bulk and slab}. The procedure for para-unitarily diagonalization reviewed in Section \ref{Colpa} generates the BV matrix $P_{\mathbf{k}}$.
The band structures displayed in Fig. \ref{fig:bands and seebeck} follow from that same procedure. The only eigenstates corresponding to bands crossing the gap are the topological edge magnons of the right and left side of the nanoribbon depicted in Fig. \ref{fig:diaghopp}.


\begin{figure*}[t]

    \subfloat[]{\includegraphics[width=0.45\textwidth]{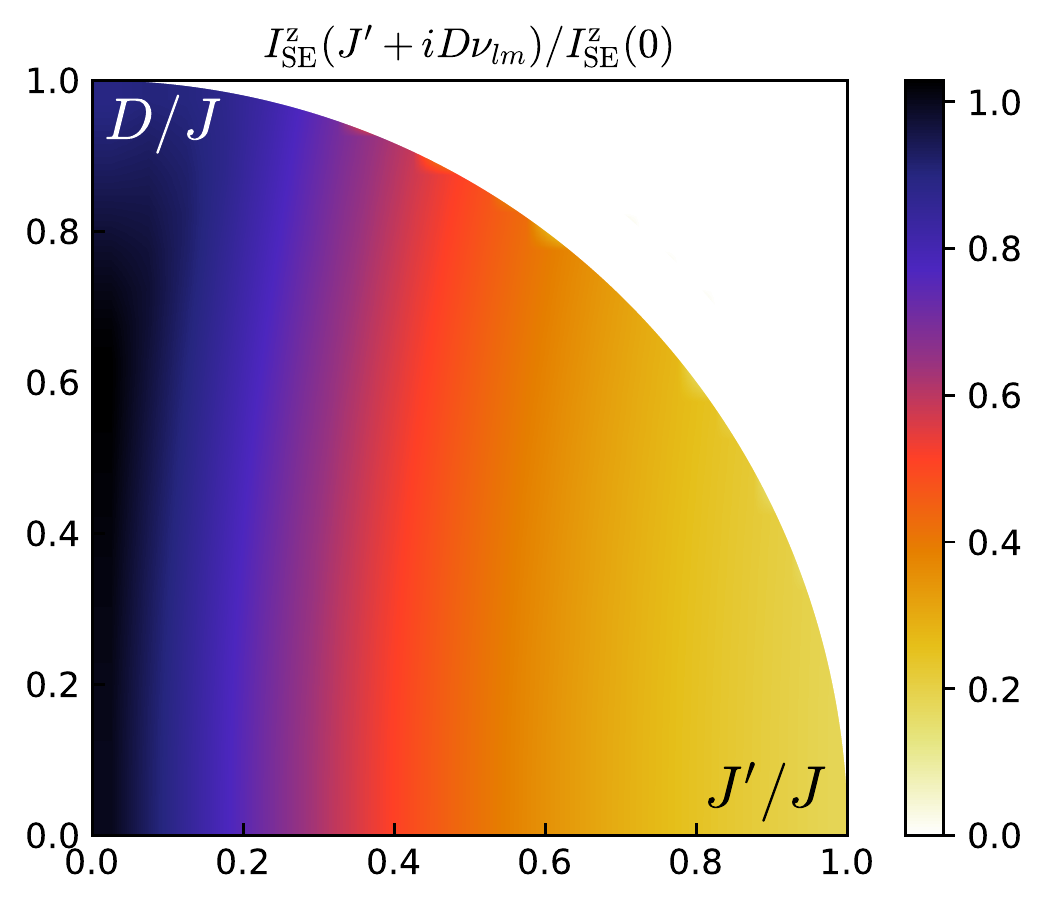}\label{fig:currentmap(a)} }%
\hfill
        \subfloat[]{\includegraphics[width=0.476\textwidth]{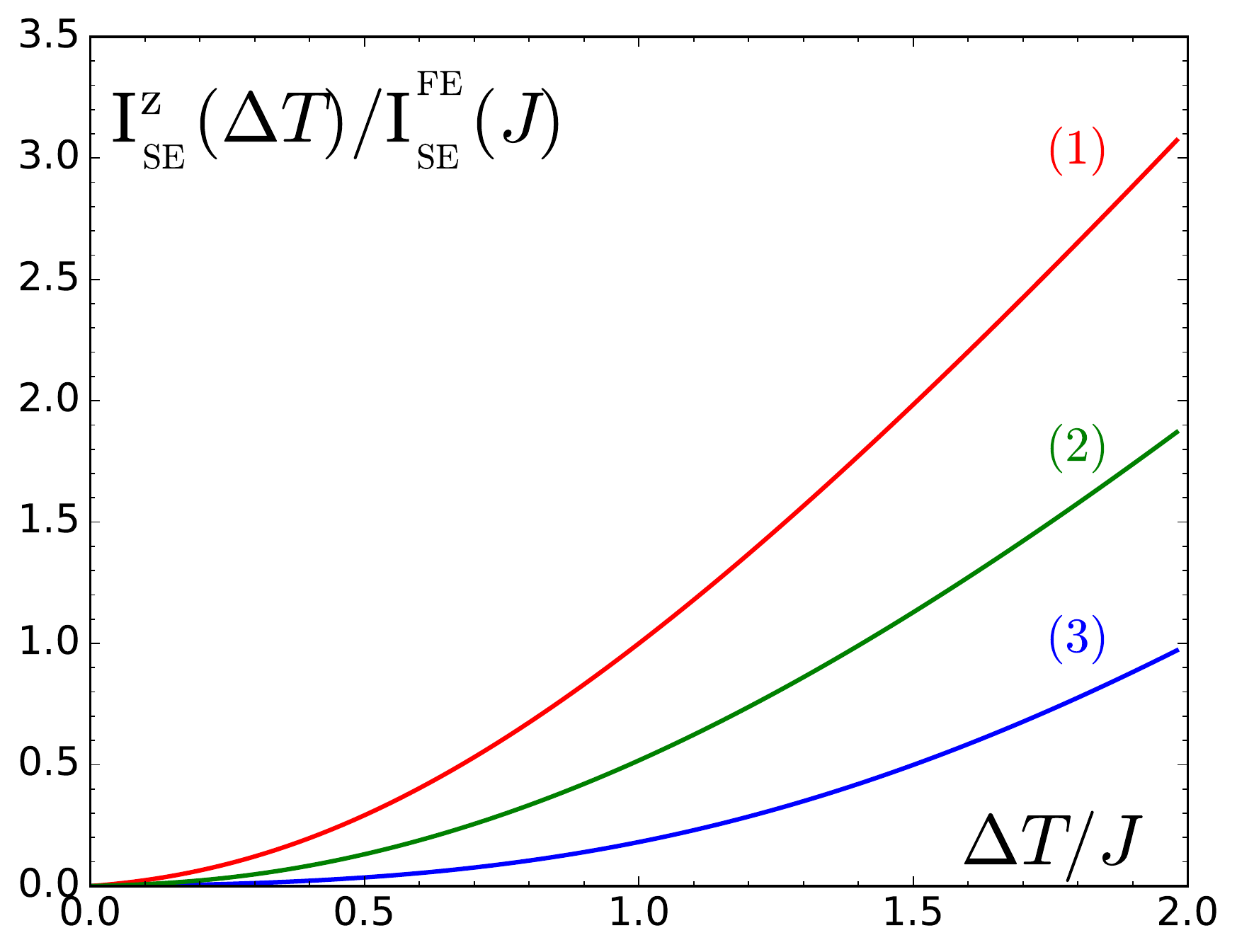}\label{fig:currentmap(b)} }%
\caption{(a) Spin current pumped through the interface, normalized by the current obtained in the simple honeycomb ferromagnet. The ribbon is composed of 15 unit cells. The color map is obtained by tuning the parameters $\mathsf{Z}=J'+iD\nu_{lm}$. The white region, for which $|\mathsf{Z}|> 1$, is not physical (nearest neighbor interaction dominates the first neighbor Heisenberg interaction) and the current is set to $0$. (b) Temperature dependence of the spin-Seebeck effect for initial equal temperature $T=0.1 J$ ($k_B=1, \hbar=1)$ and for 3 values of $(J'/J,D/J)$. (b.1) $(0,0)$, ferromagnet  $I^{\rm FE}_{\rm SE}$; (b.2) $(0.23,0.6)$; (b.3) $(0.6,0.2)$. The values are normalized by $I^{\rm FE}_{\rm SE}(\Delta T/J=1)$. }
    \label{fig:currentmap}%
\end{figure*}

\begin{figure*}
\captionsetup[subfigure]{labelformat=empty}
    \centering
    \subfloat[\centering $\mathbf{J'/J=0, D/J=0}$]{{\includegraphics[width=0.32\textwidth]{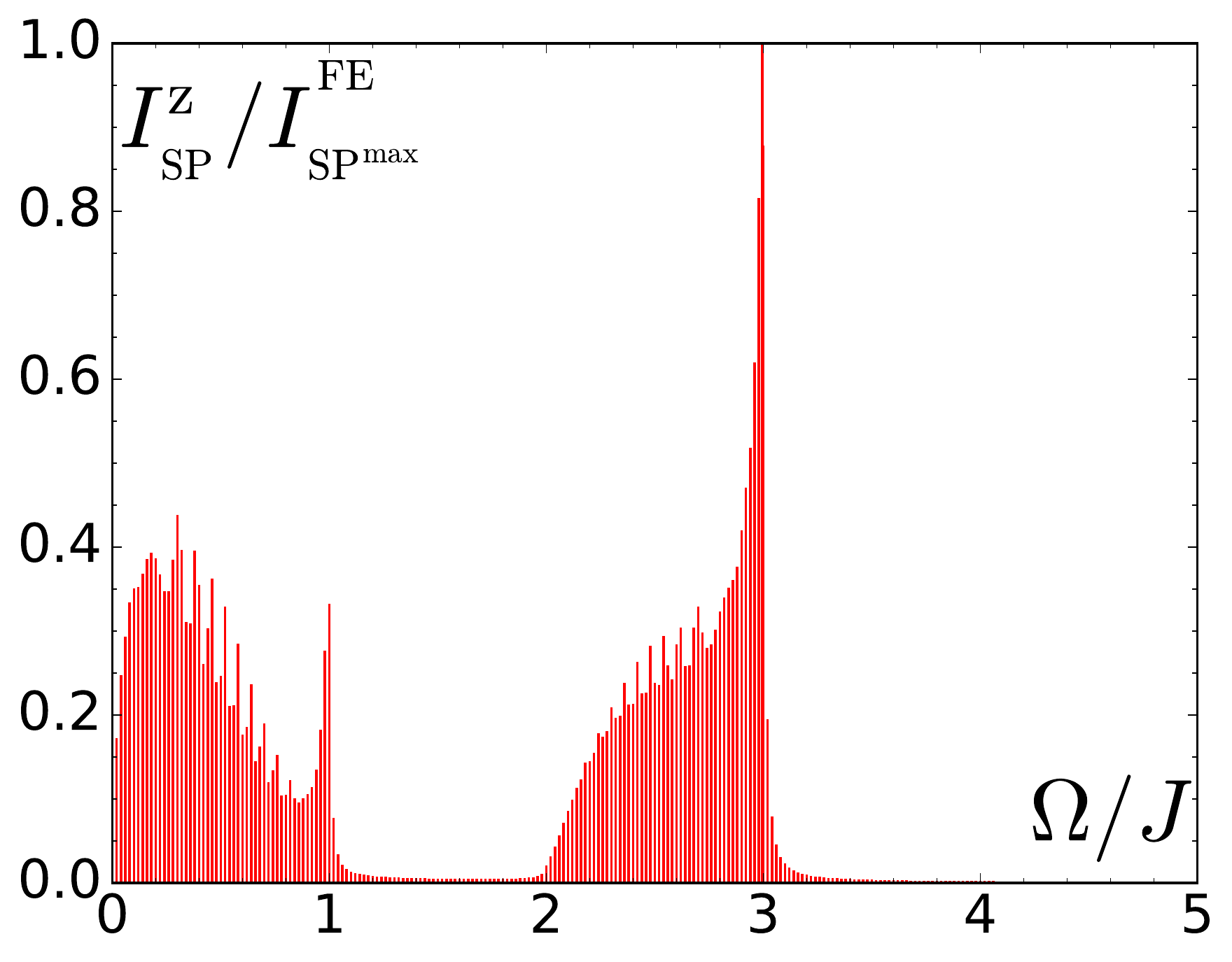} }}%
\hfill
        \subfloat[\centering $\mathbf{J'/J=0.6, D/J=0.2}$]{{\includegraphics[width=0.3\textwidth]{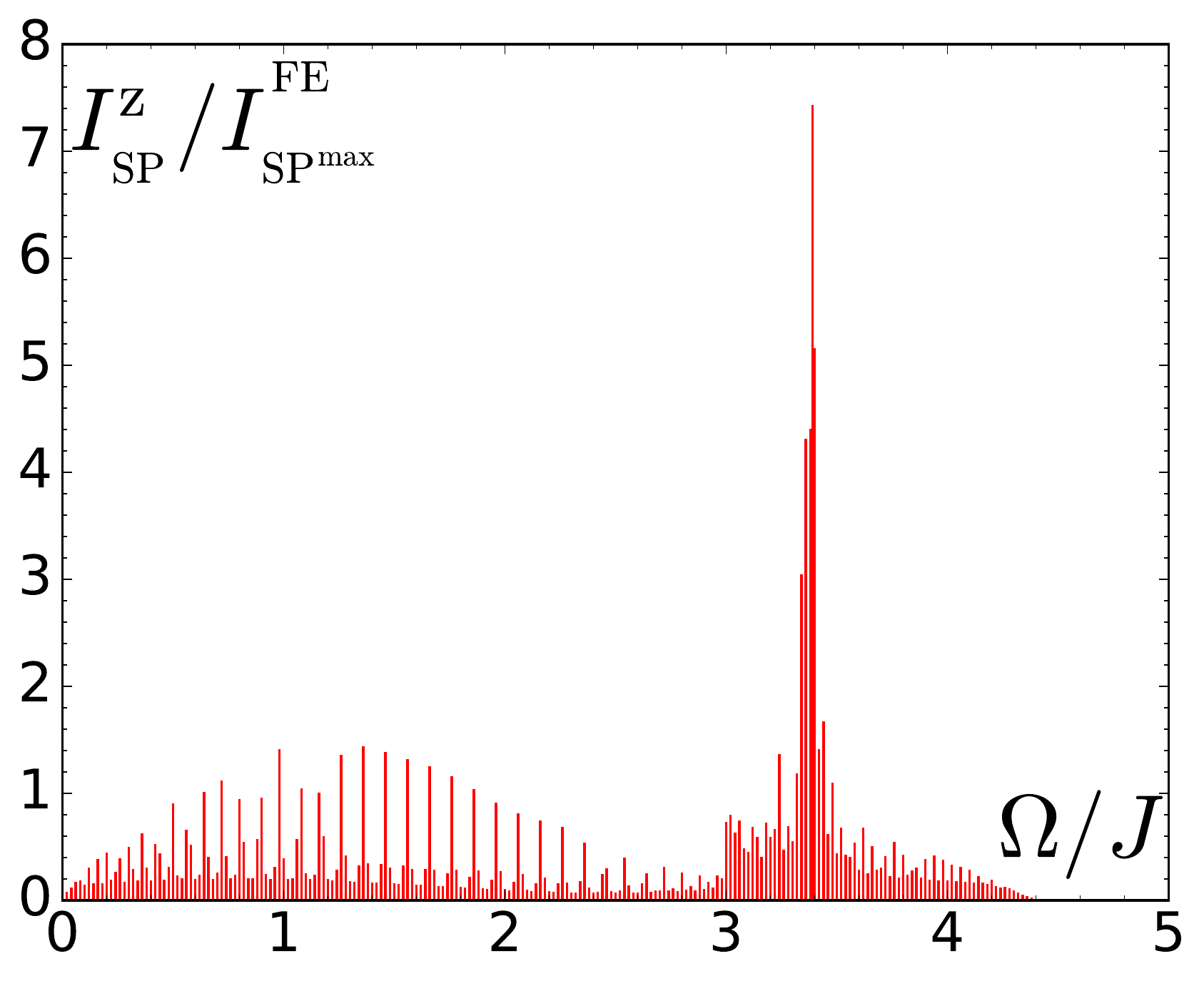} }}%
\hfill
    \subfloat[\centering $\mathbf{J'/J=0.23, D/J=0.6}$]{{\includegraphics[width=0.31\textwidth]{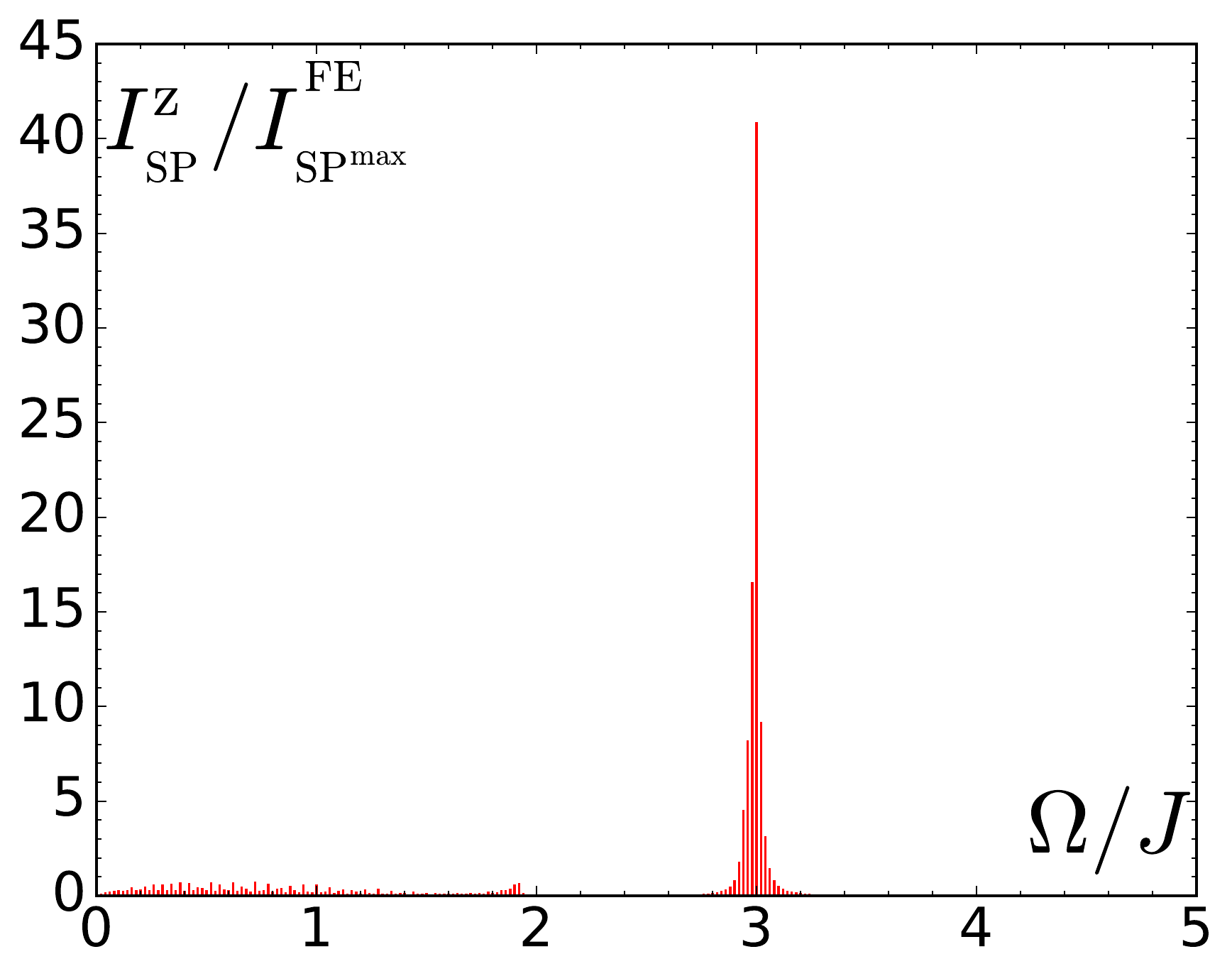} }}%
    
\flushleft
    \subfloat[\centering $\mathbf{J'/J=0, D/J=0}$]{{\includegraphics[width=0.32\textwidth]{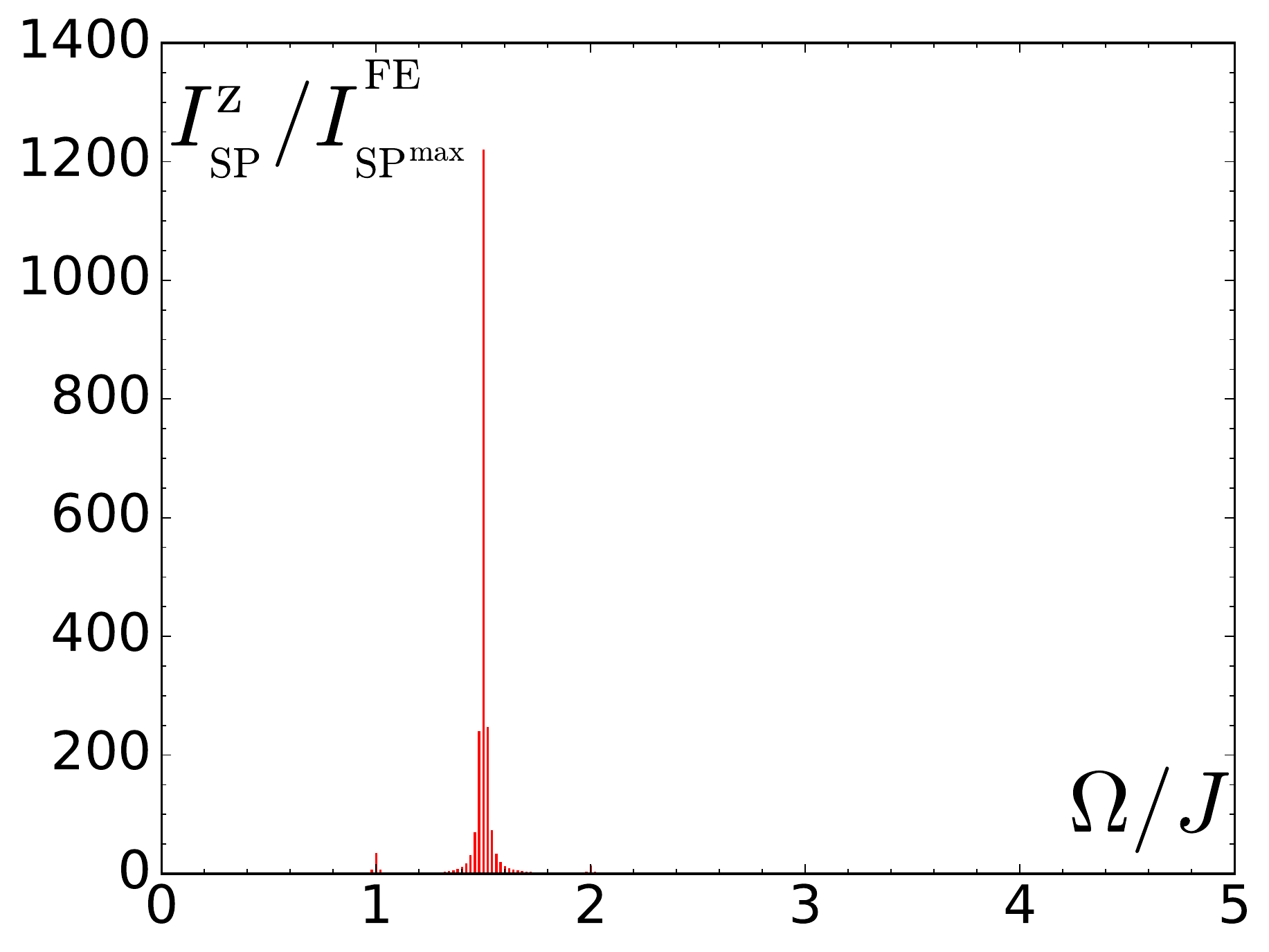}}}%
    \hfill
    \subfloat[\centering $\mathbf{J'/J=0.6, D/J=0.2}$]{{\includegraphics[width=0.32\textwidth]{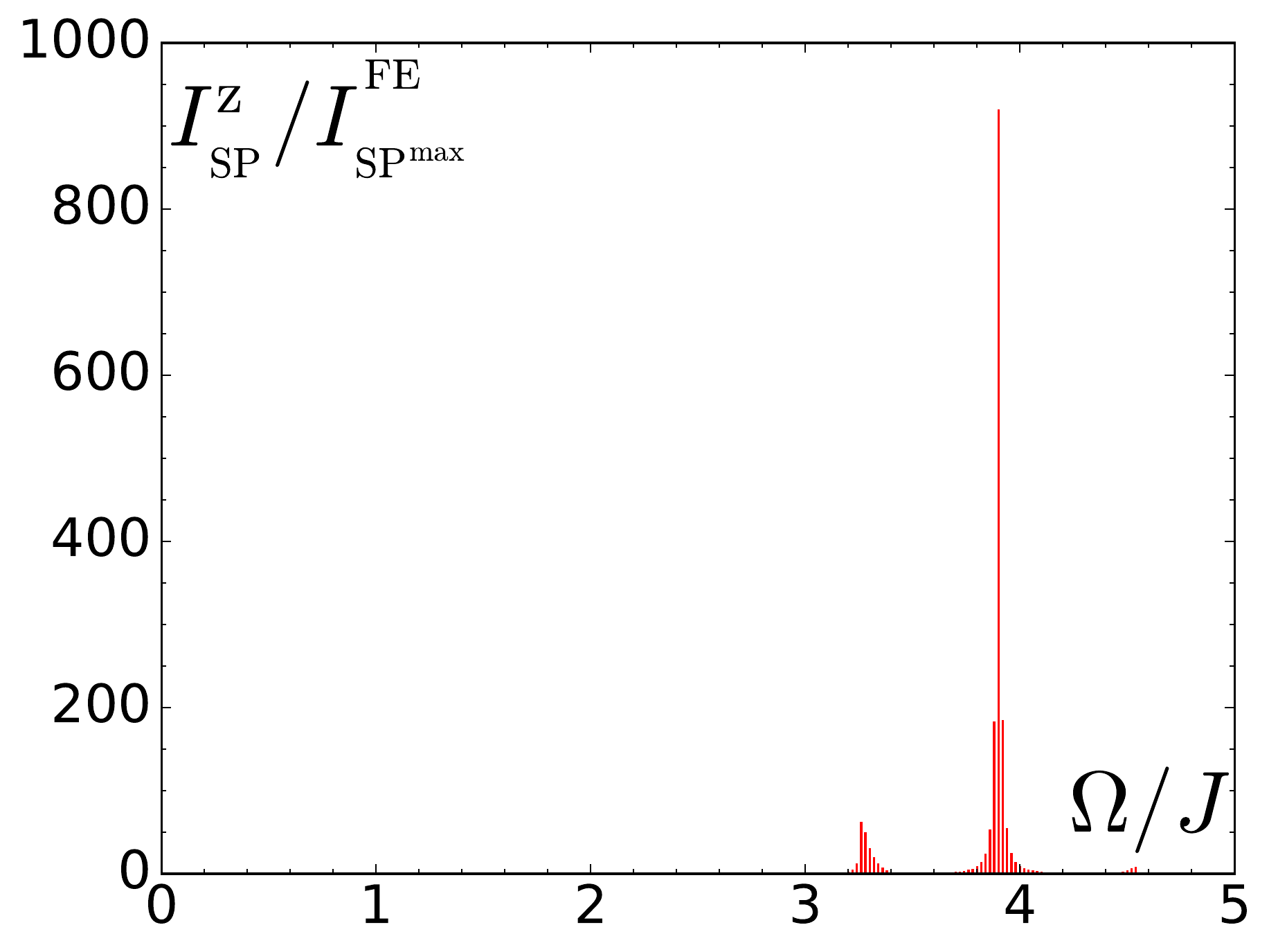} }}%
\hfill
    \subfloat[\centering $\mathbf{J'/J=0.23, D/J=0.6}$]{{\includegraphics[width=0.3\textwidth]{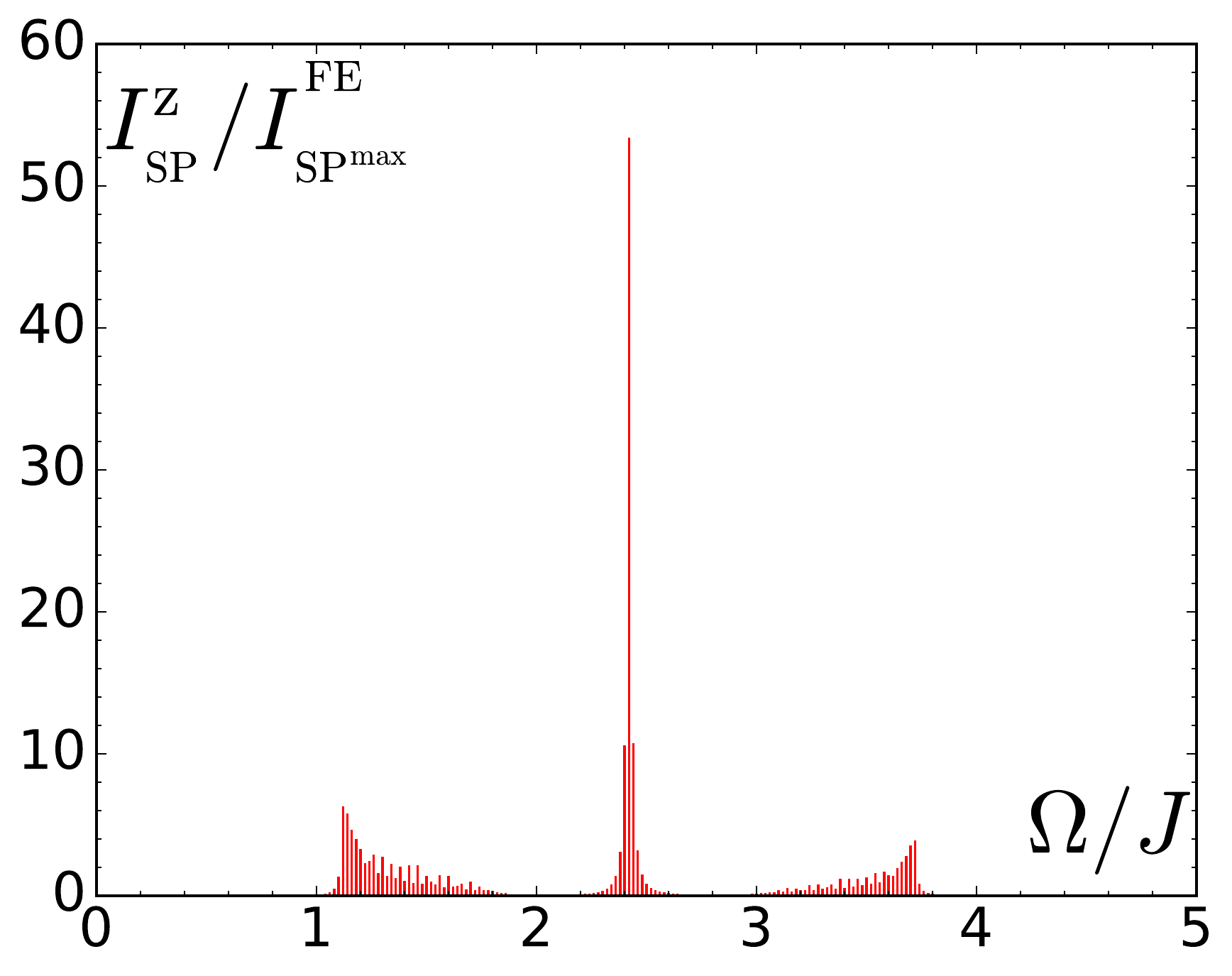} }}%
    
    \caption{Frequency contributions to the spin-current generated by the FMR drive; values are normalized by the spin-current maximal pumped in the ferromagnetic case $I^{\rm FE}_{\rm SP, max}=I^z_{\rm SP,max}(J'=D=0,k_y=0,\Omega/J=3)$. Upper line: momentum excited $k_y=0$; lower line: $k_y=\pi/a \sqrt{3}$. The ribbon is composed of 50 unit cells.}
    \label{fig:pump}%
\end{figure*}

\subsection{Spin Seebeck effect}\label{currentsection}

We analyze the spin current pumped through the interface by thermal magnons. The Hamiltonian of Eq. (\ref{eq: BDG_Owerre}) is not diagonal in space, so that Eq. (\ref{eq:ferro_diagonal_in_space}) does not hold. Then, Eq. (\ref{landauercurr1}), $\chi_i=\chi$ and the parity of $G^R$ in momentum ($\mathcal{E}_\mathbf{k}=\mathcal{E}_\mathbf{-k}$) yield
\begin{equation}
\begin{split}
\langle\hat{I}^z\rangle=A\int \frac{d\omega}{2\pi}\sum_{\mathbf{q},\mathbf{k}}&\sum_{i,j} \mathsf{M}^z_{i,j}({\mathbf{k}})  \text{Im}\chi^R(\mathbf{q},\omega)\text{Im} G^{R}_{[j]}(\mathbf{k},\omega)\\& \times \left[n_B(\omega,T_{\rm MI}) -n_B(\omega,T_{\rm NM}) \right],
 \end{split}
\end{equation}
with $A=\hbar|J|^2/2$ and $i$ runs over the two sites at the interface (see Fig. \ref{fig:diaghopp}). In this collinear setting, the role of $\mathsf{M}^z(\mathbf{k})$ is to redistribute the contribution of each magnon eigenmodes based on their Bogoliubov coefficients.

The results for the current generated by the Spin-Seebeck effect are displayed in Fig. \ref{fig:currentmap} and can be analyzed in parallel to Fig. \ref{fig:bands and seebeck}. For completeness the temperature dependence of the Spin Seebeck effect is displayed in Fig. \ref{fig:currentmap(b)}. This behavior is identical to the analysis of \cite{Matsuo2018supp}. The values of Fig. \ref{fig:currentmap(a)} are normalized by the spin current of the Heisenberg ferromagnet case, which is at the origin of the quadrant $D=J'=0$. We observe two behaviors depending on the direction in which $\mathsf{Z}$ is set. For large $J'$ the current is affected negatively. As a matter of fact, the spin-Seebeck effect does not discriminate between bulk and topological edge magnons. The main contribution to the spin current comes from the lowest energy bands even though the material is in its topological phase. Increasing $J'$ tends to relocate the gap, initially at $\epsilon/J=1.5$ for a Heisenberg ferromagnet, to $\epsilon/J>3$, and decreasing at the same time the contribution of bulk magnons. This can be rephrased as follows: interfacial magnetic moments rely on the coupling constant $J_\mathbf{qk}$ to pump but as $J'$ increases, they need to compete with the bulk interaction of the magnonic material. Furthermore, the energy gap being reduced, the specific contribution from the topological magnons would be very hard to detect.

In the low $J'$ region, the pumping magnitude is not affected much by the value of $D$. This behavior matches the opening of a large gap in the low energies $\epsilon/J<3$. The frequency spectrum of the current on Fig. \ref{fig:bands and seebeck} does however not show any specific contributions from topological magnons.

The spin-Seebeck effect study does not support any central role of topological edge magnons and cannot be used as a probe. Even though the low $J'$ regime seems favorable for pumping despite a steady magnitude at constant $J'$, the technological applications of this type of materials require a selection of edges as a controlled pumping channel.

\subsection{Spin pumping}

In the case of collinear ferromagnets, modeling an external radiation can be done by adding a term $ V\neq0 $ to the Hamiltonian, with
\begin{equation}
\begin{split}
V=&\sum_i-h_{ac}^+(t)S^+_i -h_{ac}^-(t)S^-_i.
\end{split}
\end{equation}
Here $h_{ac}^\pm (t)=\frac{\hbar\gamma h_{ac}}{\sqrt{2}}\sqrt{SN}e^{\mp\Omega t}$, $h_{ac}$, $\Omega$ being the amplitude and the frequency
of the radiated field, and $\gamma$ is the gyromagnetic ratio. Using a Holstein-Primakoff transformation, we obtain $V=\mathbf{h}(t)\mathbf{b}_{\mathbf{k}=0}$
with $\mathbf{h}(t)=(h_{ac}^-(t),..., h_{ac}^+(t),...)$ being a $2N$ row vector. Using the BV transformation, it translates into
\begin{equation}
\begin{split}
V=&\mathbf{h}'_0(t)\mathbf{a}_0,
\end{split}
\end{equation}
with $\mathbf{h}'_{\mathbf{k}}(t)=\mathbf{h}(t)P_\mathbf{k}$. This perturbation brings a correction to the lesser component of the dynamical spin susceptibility of the insulator, but does not change its retarded and advanced components. Treating $V$ pertubatively to second order yields \begin{equation}\label{landauercurr3}
\begin{split}
\delta G^<_i (\mathbf{k},\omega)&=G^R_i(\mathbf{k},\omega)\Sigma_i(\mathbf{k},\omega)G^A_i (\mathbf{k},\omega),\\
\Sigma_i(\mathbf{k},\omega)&=-\delta_{\mathbf{k},0}i\hbar^{-1}\int dt \langle \mathbf{h}'_{\mathbf{k},i}(t) \mathbf{h}'_{\mathbf{k},i+N}(0)\rangle e^{i\omega t},
\end{split}
\end{equation}
where $\mathbf{h}'_{\mathbf{k},i}\equiv (\mathbf{h}'_\mathbf{k})_i$, $1<i<N$. For consistency with previous work \cite{Kato2019,Matsuo2018}, we restore ferromagnetic damping in Eq. (\ref{eq:green function of magnons}) by the change $\vartheta \rightarrow \alpha \omega$. From Eq.~(\ref{delta fMI}), it follows that
\begin{equation}\label{landauercurr4}
\begin{split}
\delta f^{\rm MI}_i (\mathbf{k},\omega)=\frac{2\pi N S \gamma^2 }{\alpha\omega}\mathbf{h}'_{\mathbf{k},i}(0) \mathbf{h}'_{\mathbf{k},i+N}(0)\delta_{\mathbf{k},0}\delta(\omega-\Omega),
\end{split}
\end{equation}
and the current is then written
\begin{equation}\label{pumping k0}
\begin{split}
\langle\hat{I}^\alpha\rangle=& AN^2 g(\Omega)\text{Im}\chi^R_{loc}(\Omega),\\
g(\Omega)=&(S \gamma)^2\sum_{j} \frac{ \mathbf{h}'_{0,[j]}(0) \mathbf{h}'_{0,[j]+N}(0)}{(\Omega-\epsilon_{0,[j]})^2+\alpha^2\Omega^2},
 \end{split}
\end{equation}
with $\chi^R_{loc}(\Omega)=\sum_{\mathbf{q}}\chi^R(\mathbf{q},\Omega)$. We hence understand why the spin pumping setup is better suited for a control over edge magnons: by choosing $\Omega$ is the bulk gap, one mainly obtains the contribution of edge modes. An analysis over $\Omega$ then determines whether edge magnons indeed pump better into the metal than bulk magnons. 

In Fig. \ref{fig:pump}, we propose a two case study for the same values of $(J',D)$ as in Fig. \ref{fig:bands and seebeck}. The upper panels of Fig. \ref{fig:pump} uses Eq. (\ref{pumping k0}), at $k_y=0$, which is experimentally accessible. The lower panels assume that we possess an experimental procedure to excite the modes at a momentum $k_y \neq 0$. The results are normalized by the maximum pumped current of the Heisenberg ferromagnet $J'=D=0$.

For $k_y=0$, topological magnons are impossible to probe. The well defined extrema of Fig. \ref{fig:pump} correspond to a concentration of bands, as can be seen in Fig. \ref{fig:bands and seebeck}. More specifically, we verify that the concentration of bands at $k_y=0$ is a peculiarity occuring at $D/J=0.23$ and that does not depend on the value of $J'$. This coherent concentration could be interesting if one wishes to pump at defined energy ($\Omega/J=3$).
For $k_y=\pi/a\sqrt{3}$, which is the value of topological band crossing, we use
\begin{equation}\label{pumping k0}
\begin{split}
g(\Omega)=&\evalat[\Big]{(S \gamma)^2\sum_{j} \frac{\mathbf{h}'_{\mathbf{k},[j]}(0) \mathbf{h}'_{\mathbf{k},[j]+N}(0)}{(\Omega-\epsilon_{\mathbf{k},j})^2+\alpha^2\Omega^2}}{k_y=\pi/a\sqrt{3}}.
\end{split}
\end{equation}
We observe a net contribution of topological magnons due to their concentration at the edges of the magnet. The case $J'=D=0$ displays the greatest magnitude of current pumped but does not correspond to topological magnons. In contrast, at finite $J'$, the large peak is entirely due to topological magnons, demonstrating that interfacial spin pumping could be a handful approach to probe these exotic states. We emphasize that the grand challenge of the proposed approach is the ability to excite spin waves with specific momentum. Whereas this is possible in the case of dipolar magnons that possess a long wavelength (typically 100 nm) \cite{Kruglyak2010,Chumak2015}, it is much more challenging for exchange magnons. That being said, su-bmicrometer artificial magnonic crystals could offer interesting perspectives for the design and observation of topological edge states.

\section{Conclusion}\label{sec: conclusion}
The pumping properties of generic magnetic materials have been the center of the present work. Starting from the common setup consisting of an interface between a nonmagnetic metal and a magnet described in terms of a quadratic Bogoliubov-de-Gennes Hamiltonian, we gave a Landauer formula for the spin-current injected due to the spin-Seebeck effect and we extended the result to the spin pumping setup. As an application, we studied a topological ferromagnet on the honeycomb lattice in order to understand the role of topological edge magnons. Under temperature bias, we found that increasing the internal interaction of the magnet is not in favor of the spin injection. The specific contribution of topological edge modes is wiped out by the contribution of bulk magnons, which are ubiquitous due to their bosonic nature. The potential solution brought by the spin pumping setup requires to excite non-zero momentum modes, which remains an open experimental challenge but could offer appealing perspectives for topological magnonics. 

Another way to probe topological magnonic edge states could be to investigate spin current noise \cite{Matsuo2018}. Beyond bosonic edge states, it is not excluded that certain topological systems, such as frustrated magnets, can induce exotic pumping statistics that only the noise can clearly highlight.


\acknowledgments
V.G. and A.M. acknowledge support from the Excellence Initiative of Aix-Marseille Université—A*Midex, a French ‘Investissements d’Avenir’ program.\par

\appendix
\section{Metallic spin density correlators}\label{elec}

We evaluate $\left\langle (T)  s_{q_1,i}^\delta (t_1)  s_{q_2,j}^\gamma(t_2)
\right\rangle_0$ and forget initially about the dependence of operators on the position inside their unit-cell and the time ordering. These extra constraints is for now unavailing and can be added up later without loss of validity. By definition,
\begin{widetext}
\begin{equation}
\begin{split}
\left\langle  s_{\mathbf{q}_1}^\delta (t_1)  s_{\mathbf{q}_2}^\gamma(t_2)
 \right\rangle_0= \sum_{
 \sigma_{1,2},\sigma '_{1,2}=\uparrow,\downarrow 
 }\sum_{\mathbf{p}_1,\mathbf{p}_2}&\langle c^\dagger_{\sigma_1,\mathbf{q}_1+\mathbf{p}_1}(t_1)c_{\sigma'_1,\mathbf{p}_1}(t_1) c^\dagger_{\sigma_2,\mathbf{q}_2+\mathbf{p}_2}(t_2)c_{\sigma'_2,\mathbf{p}_2}(t_2)\rangle_0 \sigma_{\sigma_1\sigma_1'}^\delta  \sigma_{\sigma_2\sigma_2'}^\gamma.
\end{split}
\end{equation}
Using Wick's theorem and diagonality of $H_{\rm NM}$ in spins and momentum (in the limit of weak impurity potential),
\begin{equation}\label{wick elec}
\begin{split}
\left\langle  s_{\mathbf{q}_1}^\delta (t_1)  s_{\mathbf{q}_2}^\gamma(t_2)
 \right\rangle_0= & \sum_{
 \sigma_{1,2}=\uparrow,\downarrow 
 }\sum_{\mathbf{p}_1,\mathbf{p}_2}\langle c^\dagger_{\sigma_1,\mathbf{q}_1+\mathbf{p}_1}(t_1)c_{\sigma_1,\mathbf{p}_1}(t_1)\rangle_0 \langle c^\dagger_{\sigma_2,\mathbf{q}_2+\mathbf{p}_2}(t_2)c_{\sigma_2,\mathbf{p}_2}(t_2)\rangle_0 \delta_{\mathbf{q}_1,0}\delta_{\mathbf{q}_2,0} \sigma_{\sigma_1\sigma_1}^\delta  \sigma_{\sigma_2\sigma_2}^\gamma\\& + \langle c^\dagger_{\sigma_1,\mathbf{q}_1+\mathbf{p}_1}(t_1)c_{\sigma_1,\mathbf{p}_2}(t_2)\rangle_0 \langle c_{\sigma_2,\mathbf{p}_1}(t_1)c^\dagger_{\sigma_2,\mathbf{q}_2+\mathbf{p}_2}(t_2)\rangle_0 \delta_{\mathbf{q}_1+\mathbf{p}_1,\mathbf{p}_2}\delta_{\mathbf{p}_1,\mathbf{q}_2+\mathbf{p}_2} \sigma_{\sigma_1\sigma_2}^\delta  \sigma_{\sigma_2\sigma_1}^\gamma .
\end{split}
\end{equation}
\end{widetext}
We focus on the first line of Eq.~\ref{wick elec}. We rewrite it in terms of translational invariant electronic Green functions for which the spin dependence vanishes by diagonality of $H_{\rm NM}$ in spins,
\begin{equation}\label{componentts}
\begin{split}
 \sum_{\mathbf{p}_1,\mathbf{p}_2}\xi^<_{\mathbf{p}_1}(0) \xi^<_{ \mathbf{p}_2}(0) \sum_{
 \sigma_{1,2}=\uparrow,\downarrow 
 }\sigma_{\sigma_1\sigma_1}^\delta \sigma_{\sigma_2\sigma_2}^\gamma.
\end{split}
\end{equation}
This contribution is non zero only for $\delta=\gamma=z$. However $\Tr{\sigma^z}$=0 so that this term is inevitably vanishing. Proceeding equivalently for the second contribution of Eq.~\ref{wick elec}, we easily prove that it is non zero for $\delta=\gamma\in\{x,y\}$ only, since $\sum_{\sigma_{1,2}=\uparrow,\downarrow}\sigma_{\sigma_1\sigma_2}^\delta  \sigma_{\sigma_2\sigma_1}^\gamma=\Tr{\sigma^\delta\sigma^\gamma}=\delta_{\delta,\gamma}$. The previous reasoning applies also for time ordered average $\left\langle T s_{q_1}^\delta (t_1)  s_{q_2}^\gamma(t_2)
 \right\rangle_0$. It is now relevant to add the index over position inside the unit cell. The corresponding correlator is
$\left\langle (T)  s_{q_1}^\delta (t_1)  s_{q_2}^\gamma(t_2)
 \right\rangle_0=\left\langle (T) s_{q_1,i}^\gamma (t_1)  s_{q_2,j}^\gamma(t_2)
 \right\rangle_0
$. Replacing $s^x=\frac{1}{2}(s^+ + s^-)$ and $s^y=\frac{1}{2i}(s^+ - s^-)$, it boils down to $
\left\langle (T) s_{q_1,i}^\pm(t_1)  s_{q_2,j}^\mp(t_2)
 \right\rangle_0 $ which is diagonal in momentum. This term indicates spin mixing from site $i$ to site $j$, which we have avoided by neglecting spin-orbit coupling and non-collinear magnetism in the electronic system. Therefore this average is non-zero only when $i=j$. Using the restriction $\gamma\in \{ x,y\}$, the average appearing in the real time representation of Eq.~(\ref{generalcurr}) is expressed ($t_1=t_2=0$)
\begin{equation}
\begin{split}
   \left\langle  s_{\mathbf{q},i}^\gamma (t)  s_{\mathbf{q}',i}^\gamma(0) \right\rangle_0=&\frac{1}{4}\big [ \langle s_{\mathbf{q},i}^+ (t)  s_{\mathbf{q}',i}^-(0)  \rangle_0  \\&+\langle s_{\mathbf{q}',i}^- (t)  s_{\mathbf{q},i}^+(0)  \rangle_0\big ] \delta_{\mathbf{q},\mathbf{q}'},\\
   =&\frac{-i\hbar}{4}\left[\chi_i^>(\mathbf{q},t)+\chi^<_i(\mathbf{q},-t)\right]\delta_{\mathbf{q},\mathbf{q}'}.
\end{split}
\end{equation}
Equivalently for the time ordered terms of the real time representation of Eq.~(\ref{generalcurr}),
\begin{equation}
\begin{split}
   \left\langle T  s_\mathbf{q}^\gamma (0)  s_{\mathbf{q}'}^\gamma(t) \right\rangle_0= &\frac{1}{4}[\langle T s_{\mathbf{q}}^+ (0)  s_{\mathbf{q}'}^-(t)  \rangle_0\\&+ \langle T s_{\mathbf{q}'}^- (0)  s_{\mathbf{q}}^+(t)  \rangle_0] \delta_{\mathbf{q},\mathbf{q}'},\\
   =&\frac{-i\hbar}{4}\left[\chi^{++}(\mathbf{q},-t)+\chi^{++}(\mathbf{q},t)\right]\delta_{\mathbf{q},\mathbf{q}'}.
\end{split}
\end{equation}

\section{Current contributions}\label{sec: Current contributions}

We have proved that the contribution to the current can be express in terms of $s^\pm$, with $s^z$ inevitably vanishing. Using this result as well as the restriction $\gamma=\{x,y\}$, we can come back to the original real space definition of the current and cancel certain terms. We obtain
\begin{equation}
\begin{split}
    \hat{I}^\alpha=\frac{2}{\hbar}\sum_{m,i, \beta} J_{m,i}[\epsilon_{\alpha \beta x}& ( s^-_{m,i}S^\beta_{m,i} + \text{H.c}) \\ + i\epsilon_{\alpha \beta y}&( s^-_{m,i}S^\beta_{m,i} -\text{H.c})].
    \end{split}
\end{equation}
Taking the average, it can be written in momentum space as
\begin{equation}
\begin{split}
    \langle\hat{I}^\alpha\rangle &
    =\Re2\sum_{\mathbf{q} \mathbf{k}i \beta} J_{\mathbf{q}\mathbf{k}}\left[\epsilon_{\alpha \beta x} \langle  s^-_{\mathbf{q},i}S^\beta_{\mathbf{k} ,i}\rangle + \epsilon_{\alpha \beta y}  i\langle s^-_{\mathbf{q},i}S^\beta_{\mathbf{k} ,i}\rangle\right],
    \end{split}
\end{equation}
considering $J_{\mathbf{q} \mathbf{k} }=J_{\mathbf{k}  \mathbf{q}}^*$. We can now revisit the Keldysh operator expansion to $2$-nd order in the coupling. Keeping in mind that all $s^z$ terms vanish and that a $\delta$ function over position in the unit-cell appear,
\begin{equation}
\begin{split}
    \langle\hat{I}^\alpha\rangle
    =\Re&\Bigg[\frac{-2i}{\hbar}\sum_{ \mathbf{q} \mathbf{k}  i \beta } |J_{\mathbf{q}\mathbf{k} }|^2\\&\times\left[\epsilon_{\alpha \beta x} \int_c d\tau\langle T_c s^-_{\mathbf{q},i}s^x_{\mathbf{q},i}\rangle\langle T_c S^\beta_{\mathbf{k} ,i}S^x_{\mathbf{k} i}\rangle \right.\\&+ \left. i\epsilon_{\alpha \beta y}
    \int_c d\tau \langle T_c s^-_{\mathbf{q},i}s^y_{\mathbf{q},i}\rangle\langle T_c S^\beta_{\mathbf{k} ,i}S^y_{\mathbf{k} ,i}\rangle\right] \Bigg].
\end{split}
\end{equation}
Replacing $s^{x,y}$ by their ladder operator expression, and keeping the non-zero terms $\left\langle T s^\pm s^\mp \right\rangle_0$, we write this average
\begin{equation}
\begin{split}
    \langle\hat{I}^\alpha\rangle=\Re & \Bigg [ \frac{-i}{\hbar}\sum_{ \mathbf{q} \mathbf{k} i \beta } |J_{\mathbf{q}\mathbf{k} }|^2 \\&\times\left[\epsilon_{\alpha \beta x} \int_c d\tau\langle T_c s^-_{\mathbf{q},i}s^+_{\mathbf{q},i}\rangle\langle T_c S^\beta_{\mathbf{k} ,i}S^x_{\mathbf{k} ,i}\rangle \right.\\& + \left. \epsilon_{\alpha \beta y}
    \int_c d\tau \langle T_c s^-_{\mathbf{q},i}s^+_{\mathbf{q},i}\rangle\langle T_c S^\beta_{\mathbf{k} ,i}S^y_{\mathbf{k} ,i}\rangle\right]
  \Bigg ],
\end{split}
\end{equation}
from which the more compact expression Eq.~(\ref{current intermediaire}) follows.

\section{Magnon correlator}\label{MagnonApp}

We evaluate the two-point function $\left\langle (\Tilde{T}) S_{\mathbf{k},i}^\beta (0)  S_{\mathbf{k},i}^\gamma(t)
 \right\rangle_0$ where $\gamma \in \{x,y\}$. Starting from Eqs.~\ref{Smi} in momentum space and projecting this vector on either direction $\beta$ or $\delta$, we can summarize it into a matrix format \cite{Toth2015},
\begin{equation}
\begin{split}
\langle S^\beta_{\mathbf{k},i}S_{\mathbf{k},i}^\gamma \rangle & =\frac{S_i}{2}  \left\langle \begin{pmatrix}b_{\mathbf{k},i}^\dagger & b_{-\mathbf{k},i} \end{pmatrix}
\begin{pmatrix}
u_i^\beta u^{*\gamma}_i & u_i^\beta u^\gamma_i\\
u_i^{*\beta} u^{*\gamma}_i & u_i^{*\beta} u^\gamma_i
\end{pmatrix}
\begin{pmatrix}
b_{\mathbf{k},i}\\ b_{-\mathbf{k},i}^\dagger
\end{pmatrix}
\right\rangle,
\end{split}
\end{equation}
where the time ordering can be added on both sides. Since we would like to express this correlator in terms of magnon eigen-modes, we need to write a result in terms of $ \textbf{b}^{(\dagger)}_{\textbf{k}}=(P \textbf{a}_{\textbf{k}})^{(\dagger)}$ with $\textbf{b}^\dagger_\textbf{k}=(b^\dagger_{\textbf{k},1},...,b^\dagger_{\textbf{k},N},b_{\textbf{-k},1},...,b_{\textbf{-k},N})$. We do this in two steps. We first have
\begin{equation}\label{Bbetagammai}
\begin{split}
\langle S^\beta_{\mathbf{k},i}S_{\mathbf{k},i}^\gamma \rangle & =\frac{1}{2}  \left\langle \mathbf{v}_{\mathbf{k},i}^\dagger  \mathsf{B}^{\beta\gamma i }
\mathbf{v}_{\mathbf{k},i}
\right\rangle,
\end{split}
\end{equation}
with $\mathbf{v}_{\mathbf{k},i}=\begin{pmatrix}0\dots&b_{\mathbf{k},i} &0\dots& b_{-\mathbf{k},i}^\dagger&0\dots\end{pmatrix}^T$
whose two non-zero elements are in $i$-th and $(N+i)$-th entries. Furthermore $\mathsf{B}^{\beta\gamma i}$ is a matrix with only four non-zero elements,
\begin{equation}\label{equaaa}
\begin{split}
&\mathsf{B}^{\beta\gamma i }_{i,i}=S_iu_i^\beta u^{*\gamma}_i,\\
&\mathsf{B}^{\beta\gamma i}_{i,i+N}= S_i u_i^\beta u^\gamma_i,\\
&\mathsf{B}^{\beta\gamma i}_{i+N,i}=S_iu_i^{*\beta} u^{*\gamma i}_i,\\
&\mathsf{B}^{\beta\gamma i}_{i+N,i+N}=S_i  u_i^{*\beta} u^\gamma_i.
\end{split}
\end{equation}
After examination of Eq.~\ref{currentrealtime}, we can rather evaluate directly $\sum_{ \beta \gamma }\epsilon_{\alpha \beta \gamma}\langle S^\beta S^\gamma \rangle$. Noticing that $\mathbf{v}^{(\dagger)}$~, in Eq. (\ref{Bbetagammai}), does not depend on $\beta$ and $\gamma$, and using the identity $\sum_{ \beta \gamma }\epsilon_{\alpha \beta \gamma} \mathsf{B}^{\beta \gamma i}=\frac{1}{2}\sum_{ \beta \gamma }\left[ \epsilon_{\alpha \beta \gamma} \mathsf{B}^{\beta \gamma i}+\epsilon_{\alpha \gamma \beta} \mathsf{B}^{\gamma \beta i} \right]$, as well as $\epsilon_{\alpha\beta\gamma}=-\epsilon_{\alpha\gamma\beta}$,  the sum over spin-polarization can be compactly expressed
\begin{equation}
\begin{split}
\sum_{ \beta \gamma }\epsilon_{\alpha \beta \gamma}\langle S^\beta_{k,i}S_{k,i}^\gamma \rangle & =\frac{i}{4}\sum_{ \beta \gamma }\epsilon_{\alpha \beta \gamma}  \left\langle \mathbf{v}_{\mathbf{k},i}^\dagger \mathsf{C}^{\beta\gamma i }
\mathbf{v}_{\mathbf{k},i}
\right\rangle,
\end{split}
\end{equation}
with $\mathsf{C}^{\beta \gamma i}=\left[ \mathsf{B}^{\beta \gamma i}-\mathsf{B}^{\gamma \beta i}\right]$ a $2N\times2N$ diagonal matrix with only two non-zero elements,
\begin{equation}
\begin{split}
&\mathsf{C}^{\beta\gamma i}_{i,i}=2S_i  \Im u_i^\beta u^{*\gamma}_i, \\
&\mathsf{C}^{\beta\gamma i}_{i+N,i+N}=2S_i  \Im  u_i^{*\beta} u^\gamma_i=-\mathsf{C}^{\beta\gamma i}_{i,i}.
\end{split}
\end{equation}
The diagonality allows us to replace the right (left) vector by $\mathbf{b}^{(\dagger)}_\mathbf{k}$ without loss of validity,
\begin{equation}
\begin{split}
\sum_{ \beta \gamma }\epsilon_{\alpha \beta \gamma}\langle S^\beta_{\mathbf{k},i}S_{\mathbf{k},i}^\gamma \rangle & =\frac{i}{4}\sum_{ \beta \gamma }\epsilon_{\alpha \beta \gamma}  \left\langle \mathbf{b}^\dagger_{\mathbf{k}} \mathsf{C}^{\beta \gamma i}
\mathbf{b}_{\mathbf{k}}
\right\rangle.
\end{split}
\end{equation}
We can hence make use of the BV transform defined previously as $ \textbf{b}^{(\dagger)}_{\textbf{k}}=(P \textbf{a}_{\textbf{k}})^{(\dagger)}$ and write
\begin{equation}\label{bogoneeded}
\begin{split}
\sum_{ \beta \gamma }\epsilon_{\alpha \beta \gamma}\langle S^\beta_{\mathbf{k},i}S_{\mathbf{k},i}^\gamma \rangle & =\frac{i}{4}  \left\langle \mathbf{a}^\dagger_{\mathbf{k}} P^\dagger_{\mathbf{k}}\Upsilon^{\alpha,i} P_{\mathbf{k}}
\mathbf{a}_{\mathbf{k}}
\right\rangle,
\end{split}
\end{equation}
with $\Upsilon^{\alpha,i}=\sum_{ \beta \gamma }\epsilon_{\alpha \beta \gamma}\mathsf{C}^{\beta \gamma i}$. This sum of two-point functions is only non zero for diagonal elements since $\textbf{a}_{\textbf{k}}$ is a vector of eigenmodes. Hence,
\begin{equation}
\begin{split}
\sum_{ \beta \gamma }\epsilon_{\alpha \beta \gamma}\langle(\Tilde{T}) S^\beta_{\mathbf{k},i}(0)S_{\mathbf{k},i}^\gamma(t) \rangle & =\frac{i}{4}\sum_{j}(P^\dagger_{\mathbf{k}}\Upsilon^{\alpha,i}P_{\mathbf{k}})_{jj} \\& \times \left\langle(\Tilde{T}) (\mathbf{a}^\dagger_{\mathbf{k}})_j(0) 
(\mathbf{a}_{\mathbf{k}})_j(t)
\right\rangle.
\end{split}
\end{equation}
This can be further evaluated with Sec.~\ref{sec: green mag}.

\section{Honeycomb topological magnon insulator: Bulk and nanoribbon system}\label{Appendix: Owerre Bulk and slab}

The nearest and next nearest neighbor vectors are defined as
\begin{equation}
\begin{split}
\delta_1 &= (1,\sqrt{3})/2, \quad \delta_2 =  (1, -\sqrt{3} )/2, \quad \delta_3= -( 1, 0),\\
\rho_1 &= -(3,\sqrt{3} )/2,\quad \rho_2 = (3 ,- \sqrt{3})/ 2 ,\quad \rho_3 = (0, \sqrt{3}).
\end{split}
\end{equation}
 We use the sub-lattice configuration to express the Hamiltonian in terms of bosonic operator of $A$ and $B$ sites, respectivly noted $a_k^{(\dagger)}$ and $b_k^{(\dagger)}$. We go to Fourier space and express all hopping terms in the direction of our previously defined nearest and next nearest neighbor vector. According to our convention and following \cite{Colpa1978} we straightforwardly obtain the BdG form of the bulk system,
\begin{equation}
\begin{split}
H=&\frac{1}{2}\sum_k\begin{pmatrix}\phi^\dagger_k&\phi_{-k}\end{pmatrix}\begin{pmatrix}\mathcal{A}_k & 0\\0&\mathcal{A}^*_{-k}\end{pmatrix}\begin{pmatrix}\phi^\dagger_k\\\phi_{-k}\end{pmatrix}
+ \text{const.}\\=&\frac{1}{2}\sum_k \Psi_k \mathcal{H}_k \Psi_k
\quad \text{, with} \\ \mathcal{A}_k=&\begin{pmatrix}\nu_0-2\nu_t\sum_i \cos(\phi-\rho_i k) & -\nu_s f(k)\\-\nu_s f^*(k)&\nu_0-2\nu_t\sum_i \cos(\phi+\rho_i k)\end{pmatrix},
\end{split}
\end{equation}
and $f(k)=\sum_i e^{i\delta_ik}$, $\phi(k)=(a_k,b_k)^T$. 

We rewrite the Hamiltonian in terms of $k_y$ in a nanoribbon structure composed of one-dimensional chains stacked together. We perform this step by looking at every possible hopping between NN and NNN and by keeping track of the phase $\phi$ accumulated. The coupling matrix between the one-dimensional chains reads
\begin{equation}
\begin{split}
\mathcal{J}_{k_y}=\begin{pmatrix}L_{k_y} & T_{k_y} & 0 & & \ldots \\T^\dagger_{k_y} & L_{k_y} & T_{k_y} & 0 & \ldots \\ 0& T_{k_y}^\dagger & L_{k_y} & T_{k_y} & \dots \\ &  &   & \ddots & \end{pmatrix}.
\end{split}
\end{equation}
$L_{k_y}$ corresponds to the interaction matrix inside a single chain of the ribbon and is a hermitian matrix with upper triangular part,
\begin{widetext}
\begin{equation}
\begin{split}
L_{k_y}=\begin{pmatrix}\nu_0-2\nu_t\cos(\sqrt{3}k_y+\phi) & -2\nu_t\cos(\frac{\sqrt{3}}{2}k_y-\phi) &-2\nu_s\cos(\frac{\sqrt{3}}{2}k_y)  &-\nu_s  \\ & \nu_0-2\nu_t\cos(\sqrt{3}k_y+\phi) &0& -2\nu_s\cos(\frac{\sqrt{3}}{2}k_y) \\ & & \nu_0-2\nu_t\cos(\sqrt{3}k_y-\phi) & -2\nu_t\cos(\frac{\sqrt{3}}{2}k_y+\phi) \\ &  &   & \nu_0-2\nu_t\cos(\sqrt{3}k_y-\phi) \end{pmatrix}.
\end{split}
\end{equation}

$T_{k_y}$ models to the hopping between two adjacent chains,
\begin{equation}
\begin{split}
T_{k_y}=\begin{pmatrix}0 & -2\nu_t\cos(\frac{\sqrt{3}}{2}k_y-\phi) &0  &0  \\ 0& 0 &0& 0 \\ 0& -\nu_s& 0 & -2\nu_t\cos(\frac{\sqrt{3}}{2}k_y+\phi) \\ 0& 0 & 0  & 0 \end{pmatrix}.
\end{split}
\end{equation}
\end{widetext}


\bibliography{BiblioMagnon}
\end{document}